\documentclass[aip,pof,preprint]{revtex4-1}
\usepackage[T1]{fontenc}
\usepackage[utf8]{inputenc}
\usepackage{verbatim}
\usepackage{amsmath}
\usepackage{graphicx}
\usepackage{hyperref}
\usepackage{psfrag}
\usepackage{stmaryrd}
\usepackage{amssymb}
\usepackage{wasysym}
\usepackage{hyperref}
\usepackage{ulem}
\usepackage[dvips,usenames,dvipsnames]{color}

\usepackage{natbib}

\begin{document}

\title{Ribbon turbulence}

\author{Antoine Venaille}
\email{antoine.venaille@ens-lyon.fr}
\affiliation{Laboratoire de Physique de l'\'Ecole Normale Sup\'erieure de Lyon, CNRS and Universit\'e de Lyon, 46 All\'ee d'Italie, 69007 Lyon, France.}
\author{Louis-Philippe Nadeau}
\affiliation{PAOC, DEAPS, MIT, 77 Massachusetts Avenue, Cambridge, MA 02139-4307, USA.}
\author{Geoffrey Vallis}
\affiliation{CEMPS, University of Exeter, Exeter, EX4 4QF, UK.}

\date{\today}

\begin{abstract}

We investigate the non-linear equilibration of a two-layer quasi-geostrophic flow in a channel forced by an imposed unstable zonal mean flow, paying particular attention to the role of bottom friction. In the limit of low bottom friction, classical theory of geostrophic turbulence predicts an inverse cascade of kinetic energy in the horizontal with condensation at the domain scale and barotropization on the vertical. By contrast, in the limit of large bottom friction, the flow is dominated by ribbons of high kinetic energy in the upper layer. These ribbons correspond to meandering jets separating regions of homogenized potential vorticity. We interpret these result by taking advantage of the peculiar conservation laws satisfied by this system:  the dynamics can be recast in such a way that the imposed mean flow appears as an initial source of potential vorticity levels in the upper layer. The initial baroclinic instability leads to a turbulent flow that stirs this potential vorticity field while conserving the global distribution of potential vorticity levels.   Statistical mechanical theory of the 1-1/2 layer quasi-geostrophic model predict the formation of two regions of homogenized potential vorticity separated by a minimal interface. We show that the  dynamics of the ribbons results from a competition between a tendency to reach this equilibrium state, and baroclinic instability that induces meanders of the interface. These meanders intermittently break and induce potential vorticity mixing, but the interface remains sharp throughout the flow evolution. We  show that for some parameter regimes, the ribbons act as a mixing barrier which prevent relaxation toward equilibrium, favouring the emergence of multiple zonal jets.

\end{abstract}

\maketitle

\section{Introduction}

A striking property of observed oceanic mesoscale turbulence (from 10 to 1000 km) is the ubiquity of jets with a typical width of order the internal Rossby radius of deformation,  $R$.  In quasi-geostrophic theory $R = NH/f$ where $N$ is the buoyancy frequency, $H$ is a vertical scale and $f$ is the Coriolis parameter, and in the ocean  $R \sim 50$ km.  These jets are robust coherent structures but with high variability characterized by strong meanders --- as, for example the case of the Gulf-Stream or the Kuroshio.  Sometimes these meanders break into an isolated vortex, in which case the jets are curled into rings that literally fill the oceans.  What set the strength, the horizontal size and the vertical structure of mesoscale eddies is a longstanding problem in physical oceanography.  Here we address this question in a two-layer quasi-geostrophic model, with a particular focus on the role of bottom friction.  We consider the equilibration of an initial perturbation  in a channel with an imposed constant vertical shear $U$ in the zonal (eastward) direction. This model might be considered as one of the elementary building blocks of a hierarchy of more complex models that describe  oceanic or atmospheric turbulence~\citep{held2005gap,ColindeVerdiere09}. One motivation for this model is that the main source of energy for these turbulent flows comes from baroclinic instability that releases part of the huge potential energy reservoir set at large scale by wind forcing at the surface of the oceans or solar heating in the atmosphere~\citep{VallisBook}.

Bottom friction is the main sink of kinetic energy and without it there will be no nonlinear equilibration, so it is important to fully understand its role. A crude but effective model of that bottom friction, based on Ekman-layer dynamics, is simply linear drag with coefficient $r$. Given this,  the two-layer model has two important nondimensional parameters: the ratio $R/L_y$, with $L_y$ the width of the channel, and the ratio  $rR/U$ which is a measure of the bottom friction time scale to an inertial time scale based on the Rossby radius of deformation. There are other important parameters if the Coriolis parameter is allowed to vary but these are not our particular concern here.

In the low bottom friction limit, classical  arguments based on cascade phenomenology predict an inverse cascade of kinetic energy in the horizontal with a  concomitant  tendency toward barotropization on the vertical, i.e. the emergence of a depth independent flow~\citep{charney1971,rhines1979,Salmon_1998_Book}.  In a closed finite-sized domain, the inverse energy cascade on the horizontal leads to condensation of the eddies at the domain scale. The  intermediate regime ($rR/U\sim1$) has been studied by \citet{ThompsonYoung06JPO}, using vortex gas kinetics since the flow is made in that case of a multitude of isolated vortices or dipoles. The high bottom friction limit has been studied numerically by \citet{ArbicRibbon}~\citep{arbicflierl2004}, who also proposed scaling arguments for the vertical structure of the flow.  They observed the spontaneous formation of coherent jets  in the upper layer. The typical width of these jets was given by the Rossby radius of deformation of the upper layer. \citet{ArbicRibbon} noticed that these coherent jets looked like  localized, thin and elongated ribbons of high kinetic energy regions. These ribbons were reported to interact together in a seemingly erratic way through  meandering, pinching, coalescence and splitting of the regions separating them.  Accordingly, the high bottom friction regime  will be referred to in the following as ``ribbon turbulence''. 

The numerical results of \citet{ThompsonYoung06JPO,arbicflierl2004} were all performed in a doubly periodic domain and one novelty of our work is to consider a channel geometry. A particular advantage of the the channel geometry is that, with a proper re-definition of the potential vorticity, the dynamics in the upper layer can be recast in the form of an advection equation for the potential vorticity field, without sources or sinks, whereas in a doubly-periodic the beta term associated with the imposed mean flow must be subtracted off in order to avoid a potential vorticity discontinuity at the boundary.  We will discuss the physical consequences of these conservations laws in the low bottom friction limit and in the high bottom friction limit. This will allow us to revisit the  barotropization  process in the weak bottom friction limit, and the emergence of ribbons in the high bottom friction limit. In particular, we will interpret the emergence of ribbons as a tendency to reach a statistical equilibrium state. Statistical mechanical theory provides predictions for the self-organization properties of two-dimensional and quasi-geostrophic flows, and was initially proposed by \citet{Miller_1990_PRL_Meca_Stat,SommeriaRobert:1991_JFM_meca_Stat}. The theory applies to freely evolving flow (without forcing and dissipation), and explains self-organization of the flow into the most probable state as the outcome of turbulent stirring, and allows to compute this most probable state. In practice, the computation of the statistical equilibria requires
the knowledge of a few key parameters as the energy and the global potential vorticity distribution as an input. 

When bottom friction is large, the two-layer quasi-geostrophic dynamics is strongly dissipated, and one might expect that any prediction of the equilibrium theory applied to this two-layer flow would fail. However,  we will argue that key features of the  equilibrated states, including the emergence of ribbons,  can be accounted for by considering equilibrium states of a 1-1/2  layer quasi-geostrophic turbulence, which amounts to assume that only the upper layer is ``active".  
%One the one hand, considering a two-layer model is essential to allow for baroclinic instability in the system. On the other hand, at lowest order, the effect of bottom friction is to set to zero the velocity in the lower layer. This is why some aspects of the observed flow properties can be understood with a 1-1/2 layer model. 
%
It has been shown previously that when the Rossby radius is small, equilibrium states of the 1-1/2 layer quasi-geostrophic model contain two regions of homogenized potential vorticity, with a thin interface between these regions~\citep{BouchetSommeriaJFM02}.  We will explain why this is relevant to describe the emergence of ribbons and provide a complementary point of view based on cascade arguments. We will  go further than the application of equilibrium statistical mechanics  in order to account for some of the dynamical aspects  of the ribbons.   In particular, we will show that the observed meanders of the ribbons cannot be explained in the framework of 1-1/2  layer quasi-geostrophic model, but must be accounted  for by the baroclinic instability of the ribbons in the framework of a two-layer quasi-geostrophic model. We will also see that  once a ribbon is formed, it may act as a mixing barrier and prevent relaxation towards the equilibrium state. For this reason, more than two regions of homogenized regions can coexist for some range of parameters. We will relate this observation to the emergence of multiple zonal jets in this flow model.  

The paper is organized as follows. The basic model is presented in section \ref{sec:barocturb} along with  a discussion of the physical consequences of existing conservation laws for the dynamics. In section \ref{sec:predic} we review existing results based on cascade arguments and statistical mechanics approach and give predictions for the flow structure at large times. These predictions are tested against numerical simulations in a section \ref{sec:test}, and we conclude in section \ref{sec:conclude}.

\section{Baroclinic turbulence in a two-layer quasi-geostrophic flow}
\label{sec:barocturb}

\subsection{Two layer quasi-geostrophic flows in a channel}

We consider a two-layer quasi-geostrophic model on a $f$-plane in a  channel periodic in the $x$ direction and of size $(L_x \times L_y)$  (Fig \ref{fig:pv_init}-a). 
The relative depth of the upper and lower layers are $\delta=H_1/H$ and $1-\delta=H_2/H$, respectively, with $H$ the total depth. Consequently, the internal Rossby radius of deformation of the upper and the lower layer are $R_1=\delta^{1/2} R$ and $R_2=(1-\delta)^{1/2}R$, respectively, with $R=(Hg')^{1/2}/ f_0$, with $g'$ the reduce gravity between the two layers, and $f_0$ the Coriolis parameter.   The dynamics is given by the advection
in each layer of the potential vorticity fields $q_1,q_2$ by a non-divergent velocity
field which can be expressed in term of a streamfunction   $\Psi_1, \Psi_2$ :
\begin{equation}
\partial_{t}q_{1}+J\left(\Psi_{1},q_{1}\right)=-A_h\nabla^6 \Psi_1,
\label{eq:dyn_pv1bis}
\end{equation}
\begin{equation}
\partial_{t}q_{2}+J\left(\Psi_{2},q_{2}\right)=-A_h\nabla^6 \Psi_2-r \nabla^2 \Psi_{2}\ ,
\label{eq:dyn_pv2bis}
\end{equation}
where  $A_h$ is a lateral bi-harmonic viscosity coefficient, $r$ is a bottom drag coefficient, $J(a,b)=\partial_{x}a\partial_{y}b-\partial_{y}a\partial_{x}b$
is the Jacobian operator. The velocity field in each layer is given by $U_{i}=-\partial_{y}$$\Psi_{i}$,
$V_{i}=\partial_{x}\Psi_{i}$, for $i=1,2$. The potential vorticity
fields are expressed as the sum of a relative vorticity term $\zeta_i=\nabla^2 \Psi_i$   and a stretching term involving the
Rossby radius of deformation $R$:
\begin{equation}
q_{1}=\nabla^2\Psi_{1}+\frac{\Psi_{2}-\Psi_{1}}{\delta R^{2}} ,\label{eq:pv1bis}\end{equation}
\begin{equation}
q_{2}=\nabla^2 \Psi_{2}+\frac{\Psi_{1}-\Psi_{2}}{\left(1-\delta\right)R^{2}} \ .\label{eq:pv2bis}
\end{equation}

These equations must be supplemented with boundary conditions. The flow is periodic in the $x$ direction, and there is no flow across the wall at the northern and the southern boundaries. This  impermeability constraints amounts to assume  that  $\Psi_{1,2}$ is a constant at the northern and the southern boundary.  Four equations are then needed to determine these constants. Two equations are given by mass conservation, which imposes the constraints 
\begin{equation}
\int_{\mathcal{D}} \mathrm{d}x\mathrm{d}y \Psi_1=\int\mathrm{d}x\mathrm{d}y \Psi_2=0 \ . \label{eq:mass_cons}
\end{equation}
Two additional equations are obtained by integrating over one line of constant latitude (constant $y$) the zonal projection (along $\mathbf{e}_x$) of the  momentum equations in each layers. Let us consider the particular case where the line of constant latitude is the southern boundary, and let us call $\Gamma_i=\int_0^{L_x}\mathrm{d} x U_i(x,0)$ the circulation along this boundary. Then the two additional equations are 
\begin{equation}
\frac{d\Gamma_1}{dt}=-A_h\left( \int_0^{L_x} \mathrm{d} x \nabla^4 U_1\bigg|_{y=0}\right) ,\quad \frac{d\Gamma_2}{dt}=-A_h \left(\int_0^{L_x} \mathrm{d} x \nabla^4 U_2\bigg|_{y=0}\right) -r\Gamma_2 ,  \label{eq:circulations_bis}
\end{equation} 
see \citet{pedlosky1982} for further details on the quasi-gesotrophic dynamics in an open channel. 

In the absence of small scale dissipation (i.e. when $A_h=0$), the dynamics is fully determined by Eq. (\ref{eq:dyn_pv1bis}-\ref{eq:dyn_pv2bis}-\ref{eq:circulations_bis}-\ref{eq:mass_cons}).   When small scale dissipation is taken into account (i.e. when $A_h\ne 0$), additional boundary conditions are required due to higher order terms appearing in Eq.  (\ref{eq:dyn_pv1bis}-\ref{eq:dyn_pv2bis}-\ref{eq:circulations_bis}) . We will consider in numerical simulations a free-slip boundary condition:  the vorticity is set to zero at the southern and northern boundary of each layer. 

\subsection{Evolution of a perturbation around an imposed mean flow}

We impose the existence of a  constant eastward flow in the upper layer with a lower layer at rest ($\overline{\Psi}_1=-Uy$, $\overline{\Psi}_2=0$).  We denote $\psi_1$ and $\psi_2$ the perturbation around this mean flow  ($\psi_i=\Psi_i-\overline{\Psi}_i$). The potential vorticity fields defined Eq. (\ref{eq:pv1bis}-\ref{eq:pv2bis}) can be written in term of this perturbed streamfunction :
\begin{equation}
q_{1}=\nabla^2 \psi_{1}+\frac{\psi_{2}-\psi_{1}}{\delta R^{2}}+\frac{U}{\delta R^{2}}y\ ,\label{eq:pv1}
\end{equation}
\begin{equation}
q_{2}=\nabla^2 \psi_{2}+\frac{\psi_{1}-\psi_{2}}{\left(1-\delta\right)R^{2}}-\frac{U}{\left(1-\delta\right)R^{2}}y\ .\label{eq:pv2}
\end{equation}
We see that the mean flow is associated with a zonal  potential vorticity gradient (an "effective beta plane" term) having an opposite sign in the upper and lower layer.  The dynamics of the perturbation is then fully described by the potential vorticity advection:
\begin{equation}
\partial_{t}q_{1}+J\left(\psi_{1}-U y,q_{1}\right)=- A_h \nabla^6 \psi_1\ ,\label{eq:dyn_pv1}\end{equation}
\begin{equation}
\partial_{t}q_{2}+J\left(\psi_{2},q_{2}\right)- A_h \nabla^6 \psi_2-r \nabla^2 \psi_{2}\ .\label{eq:dyn_pv2}
\end{equation}
When this equation is linearized around the mean flow, we recover the Philipps model for baroclinic instability on a $f$-plane, see e.g.  \citet{VallisBook}. In this configuration, the mean flow is always unstable and the most unstable mode is always  associated with an horizontal scale that scales with the internal Rossby radius of deformation, whatever the value of bottom friction.  Only the time scale for the instability changes with bottom friction. Our aim is to study the non-linear equilibration of this instability.

\subsection{Conserved quantities}

\begin{figure}[t]
\begin{center}
\includegraphics[width=\textwidth]{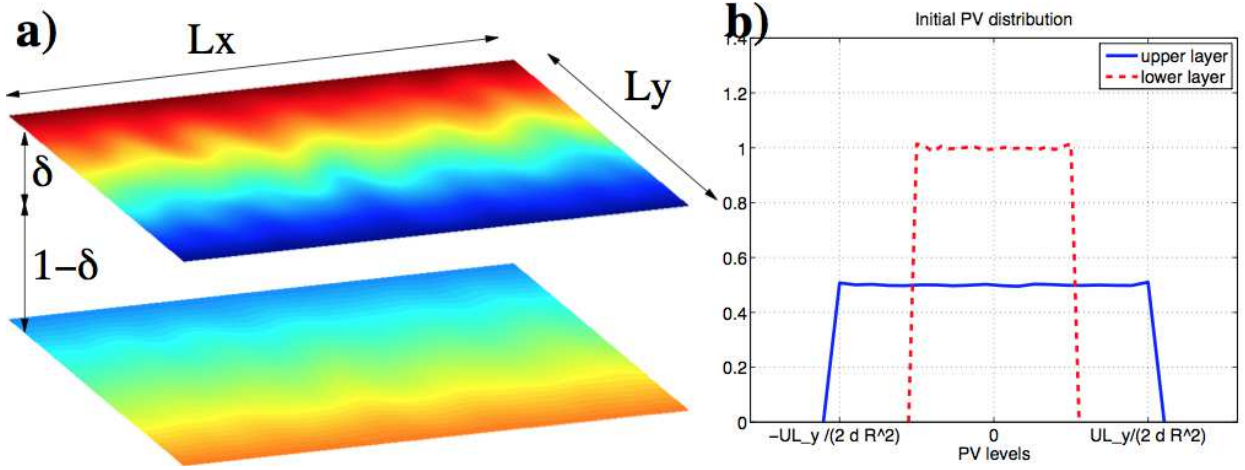}
\end{center}
\caption{Sketch of the numerical experiment. Left panel: potential vorticity field at the beginning of the simulation, when the most unstable mode start to grow, in the case without bottom friction. Right panel global distribution of potential vorticity levels at t=0, which is the same whatever the bottom friction.   \label{fig:pv_init}.}
\end{figure}

The flow model has a remarkable property: in the absence of small scale dissipation, the potential vorticity in the upper layer $q_1$ is advected without sinks nor sources. As a consequence, there is an infinite number of conserved quantities, namely the Casimir functionals $\mathcal{C}_s[q_1]=\int_{D} \mathrm{d}x\mathrm{d}y  \ s(q_1)$, where $s$ is any sufficiently smooth function, see also \citet{shepherd1988}.   An equivalent statement is that the global distribution of the potential vorticity levels in the upper layer  is conserved through the flow evolution when there is no small scale dissipation. Since the initial flow is characterized by $q_1\big|_{t=0}=Uy/(\delta R^2)$, the global distribution of fine grained potential vorticity in the upper layer is a flat distribution of potential vorticity levels between $-UL_y/(2\delta R^2)$ and  $UL_y/(2\delta R^2)$, see Fig. \ref{fig:pv_init}-b. Similarly, the global distribution of the potential vorticity in the lower layer is conserved  if both the small scale dissipation and the bottom friction are zero. In that case, given our initial potential vorticity profile, the global distribution of potential vorticity levels in the lower layer is a flat distribution between $-UL_y/(2\delta R^2)$ and  $UL_y/(2(1-\delta) R^2)$, see Fig. \ref{fig:pv_init}. If bottom friction is non zero, the potential vorticity distribution of the lower layer is not conserved, but the potential vorticity distribution of the lower layer remains bounded.  Remarkably, the presence of bottom friction does not affect conservation of the potential vorticity distribution in  the upper layer.

When there is small scale dissipation, the global distribution of potential vorticity levels is no more a conserved quantity. However, if the time scale for the relaxation of the initial condition towards a quasi-stationary state is  smaller than the typical dissipation time scale, then one expects that the conservation laws of the inviscid dynamics still plays an important role. %This is why we propose to interpret our numerical simulati on by considering these conservation laws, which will be shown to have important consequences for the equilibration of the flow, either in the limit of very large bottom friction or in the limit of very small bottom friction.  

\subsection{Energy budget}\label{sec:energy}
%R^2=g'H/f^2
%Ld^2=R^2 H1 H2 /H^2
%delta_arbic=H1/H2 =delta/(1-delta)
%Ld^2=R^2  delta (1-delta)
The  energy of the perturbation is the sum of kinetic energy in each layer and of the available potential energy:
\begin{equation}
E=KE_1 +KE_2+APE, \quad APE= \frac{1}{2}\int_{\mathcal{D}}\mathrm{d}x\mathrm{d}y\,\frac{\left(\psi_{1}-\psi_{2}\right)^{2}}{R^{2}},
\end{equation}
\begin{equation}
KE_1=\frac{\delta}{2}\int_{\mathcal{D}}\mathrm{d}x\mathrm{d}y\,\left(\nabla\psi_{1}\right)^{2},\quad KE_2= \frac{1-\delta}{2}\int_{\mathcal{D}}\mathrm{d}x\mathrm{d}y\,\ \left(\nabla\psi_{2}\right){}^{2} \ .  \label{eq:Energy}\end{equation}
In the absence of  small scale dissipation, the  temporal evolution of the energy of the perturbation  is given by 
\begin{equation}
\frac{R}{U}\frac{\mathrm{d}}{\mathrm{d}t}E= \frac{1}{R} \int_{\mathcal{D}}\mathrm{d}x\mathrm{d}y\ \psi_{1} \partial_{x}\psi_{2}-(1-\delta)\frac{rR}{U}\int_{\mathcal{D}}\mathrm{d}x\mathrm{d}y\ \left(\nabla\psi_{2}\right)^{2} \ .\label{eq:Energy-dynamics}
\end{equation}
We readily note that the parameter $rR/U$ plays a key role in the energy budget (\ref{eq:Energy-dynamics}), and that this energy budget for the perturbed flow is the same as one would obtain in the doubly periodic geometry~\citep{arbicflierl2004}.  In the channel geometry, it is  also useful to introduce the "total energy" defined as the energy of the flow that includes the perturbation and the mean flow: 
\begin{equation}
E_{tot}=KE_{1tot}+KE_{2tot}+ APE_{tot} , \quad   APE_{tot} =   \frac{1}{2}\int_{\mathcal{D}}\mathrm{d}x\mathrm{d}y\, \frac{\left(\psi_{1}-Uy-\psi_{2}\right)^{2}}{R^{2}},
\end{equation}
\begin{equation} 
 KE_{1tot}= \frac{\delta }{2}\int_{\mathcal{D}}\mathrm{d}x\mathrm{d}y\, \left(\nabla\psi_{1}-Uy\right){}^{2},  \quad  KE_{1tot}= \frac{1-\delta }{2} \int_{\mathcal{D}}\mathrm{d}x\mathrm{d}y\, \left(\nabla\psi_{2}\right){}^{2}.  \label{eq:Energy_tot}
 \end{equation}
 The temporal evolution of the total energy is given by 
 \begin{equation}
\frac{\mathrm{d}}{\mathrm{d}t}E_{tot}= -(1-\delta) r  \int_{\mathcal{D}}\mathrm{d}x\mathrm{d}y\ \left(\nabla\psi_{2}\right)^{2} \ .\label{eq:Energy-dynamics-tot}
\end{equation}
%Note that there  would be additional source terms in the right hand side in the case of a doubly periodic domain. 
 This equations for the total energy allows for a clear physical interpretation in the channel case: in the presence of bottom friction, the total energy will decay to zero.  In other words, the perturbation will  evolves toward the state  $\psi_1=-Uy$, $\psi_2=0$ which  annihilates the imposed mean flow.   We see from Eq. (\ref{eq:pv1}-\ref{eq:pv2}) that such a state corresponds to  fully homogenized potential vorticity fields $q_1=q_2=0$.  Note that this potential vorticity homogenization process does not rely on the existence of small scale dissipation, since the potential vorticity can be homogenized at a coarse grained level. The important mechanism is the filamentation process following sequences of stretching and folding of the potential vorticity field through turbulent stirring.% We also note that when this fully homogenized state is reached, the kinetic and the potential energy of the perturbed flow are the same as the kinetic and  the potential energies of the imposed mean flow.
 
 We will see in the following that the route towards complete potential vorticity homogenization strongly depends on the parameter $rR/U$. In particular, dimensional analysis predicts that the time scale for homogenization can be written on the general form 
\begin{equation}
t_{diss} \sim \frac{1}{r}F_{diss}\left( \frac{rR}{U},\frac{R}{L_y},\delta,\frac{L_x}{L_y}\right), \label{eq:t_diss_general}
\end{equation}
where the argument of the function $F_{diss}$ are the four non-dimensional parameters of the problem, assuming vanishing small scale dissipation ($A_h=0$). 
We will argue in the next section that when the domain is large with respect to the Rossby radius of deformation ($L_y\gg R $), when the upper layer is thin with respect to the total depth ($ \delta \ll 1$) and when the domain aspect ration is of order one ( $L_x \sim L_y$) the function $F_{diss}$ can be modeled by xx check if $\delta \ll 1$ is necessary xx.
\begin{equation}
F_{diss}=1 +   \frac{1-\delta}{\delta^{1/2}} \left(\frac{rR}{U}\right)^2\frac{L_y}{R} . \label{eq:t_diss_F}
\end{equation}
For that purpose, we will need to discuss the  vertical and the horizontal flow structure at large time, before complete homogenization is achieved.  
 
\section{Predictions for the flow structure at large time}
\label{sec:predic}

The aim of this section is to provide predictions for the vertical partition of the energy, and to explore consequences of this vertical structure for the self-organization of the flow on the horizontal. We first show that barotropization is expected for vanishing bottom friction. We then explain that surface intensification is expected for large bottom friction. We then use a combination of arguments based on cascade phenomenology, potential vorticity homogenization theories and equilibrium statistical mechanics in order to predict the horizontal flow structure in the large bottom friction limit and the small bottom friction limit.  
It is assumed in this section that the small scale dissipation is negligible ($A_h=0$).

\subsection{Barotropization  in the low bottom friction limit}\label{sub:barotropization}

We consider first the case with zero bottom friction ($r=0$). It will be useful to consider 
the barotropic and baroclinic modes of the two-layer model, defined as 
\begin{equation}
\psi_t=\delta \psi_1 + \left( 1 - \delta \right) \psi_2 , \quad \psi_c = \psi_1 - \psi_2  \ .\label{eq:baro-modes}
\end{equation}
The baroclinic streamfunction $\psi_c$ and the barotropic streamfunction $\psi_t$  are related to the potential vorticity through 
\begin{equation}
q_1 - q_2 = \nabla^2 \psi_c- \frac{ \psi_c}{ \delta (1-\delta ) R^2}-\frac{ U}{ \delta (1-\delta ) R^2} y\label{eq:barocline}
\end{equation}
\begin{equation}
\delta q_1 +(1-\delta)q_2  = \nabla^2 \psi_t \label{eq:barotrope}
\end{equation}
The energy of the perturbation can be decomposed into a (purely kinetic) barotropic energy and a baroclinic energy that involves both kinetic energy and potential energy:
\begin{equation}
E=KE_{t}+KE_{c}+APE_{c}, \quad APE_c = \frac{1}{2}\int_{\mathcal{D}}\mathrm{d}x\mathrm{d}y\, \frac{\psi_c^{2}}{R^{2}}. \label{define_Et}
\end{equation}
\begin{equation}
KE_t=\frac{1}{2}\int_{\mathcal{D}}\mathrm{d}x\mathrm{d}y\, \left(\nabla\psi_{t}\right)^{2},\quad KE_c= \delta \left(1-\delta\right)   \frac{1}{2}\int_{\mathcal{D}}\mathrm{d}x\mathrm{d}y\, \left(\nabla\psi_{c}\right)^{2}. \label{eq:Energy}
\end{equation}
Similarly, the total energy can be decomposed into a barotropic and a baroclininc component:
\begin{equation}
E=KE_{tot,t}+KE_{tot,c}+APE_{tot,c}, \quad APE_{tot,c} = \frac{1}{2} \int_{\mathcal{D}}\mathrm{d}x\mathrm{d}y\, \frac{\left(\psi_c-Uy\right)^{2}}{R^{2}}. \label{define_Etot}
\end{equation}
\begin{equation}
KE_{tot,t}=\frac{1}{2}\int_{\mathcal{D}}\mathrm{d}x\mathrm{d}y\, \left(\nabla(\psi_{t} -\delta U y) \right)^{2},\quad KE_{tot,c}= \delta \left(1-\delta\right)   \frac{1}{2} \int_{\mathcal{D}}\mathrm{d}x\mathrm{d}y\, \left(\nabla\left(\psi_{c}-U y\right)\right)^{2}. \label{eq:Energy_{tot}}
\end{equation}

 The initial potential vorticity fields in the upper and lower layers are respectively  $q_1^0=Uy /\delta R^2$ and $q_2^0=-U y /(1-\delta) R^2$, plus a small perturbation. When $R \ll L_y$ or $\delta \ll 1$, and when $L_x\sim L_y$,  the   initial total energy is dominated by the potential energy: $E^0_{tot}\sim APE_{tot}^0 \sim U^2L_y^4/R^2$. %We discuss in the following how the different energy contributions vary during the turbulent evolution of the flow in the absence of bottom friction.

The classical picture for two layer geostrophic turbulence predicts that the turbulent evolution of the flow leads to  barotropization~\citep{charney1971,rhines1979,Salmon_1998_Book}, i.e. to a depth independent flow   for which $E_{tot}\approx KE_{tot,t}$. In the context of freely evolving inviscid dynamics, the idea that barotropization may occur as a tendency to reach a statistical equilibrium state that takes into account dynamical invariants  has been investigated  by Refs. \citep{VenailleVallisGriffies,HerbertPRE,renaud_etal_preprint}. It was found in these studies that  barotropization may be prevented by conservation of potential vorticity levels in some cases. We provide in Appendix A  a phenomenological argument for barotropization in the case $R\ll L_y$ or $\delta\ll 1$, emphasizing the role of the conservation of potential vorticity levels, and of the total energy. In this limit, the flow dynamics  is described at lowest order by the barotropic dynamics after its initial turbulent rearrangement:
\begin{equation}
   \partial_t q_t +J(\psi_t,q_t)=0,\quad q_t=\nabla^2 \psi_t  . \label{eq:baro}
\end{equation}
%We estimate in Appendix A  the order of magnitude for the barotropic and the baroclinic energy after turbulent rearrangement  of. %We show in the following that this configuration is a very peculiar one, and 
%n other words, the initial energy dominated by available potential energy is transferred into (purely kinetic) barotropic energy through turbulent rearrangement of the potential vorticity field. 
Let us now discuss the effect of a weak friction  $rR/U \ll 1 $.  Let us call $t_{adv} = L_y/U$ the typical advection time scale for the flow over the whole domain. This can be considered as the typical time scale  for the self-organization of the turbulent dynamics following the initial instability that occurs on a time scale $t_{inst} = R/U$.  Once the flow is self-organized at the domain scale, if the flow is dominated by the barotropic mode,  we see Eq. (\ref{eq:Energy-dynamics-tot}) that the total energy should  decay exponentially with an e-folding time $t_{fric} \sim 1/(1-\delta )r$.   This  justifies the low friction limit for the function $F_{diss}$  defined Eq. (\ref{eq:t_diss_F}).

\subsection{Surface intensification in the large bottom friction limit}\label{sub:surface_intensification}

In the large bottom friction limit, if the system reaches or a quasi-stationary state,  we see from the energy budget  Eq. (\ref{eq:Energy-dynamics}) for the perturbed flow  that the friction term $(1-\delta)(rR/U)\int_{\mathcal{D}}\mathrm{d}x\mathrm{d}y\ \left(\nabla\psi_{2}\right)^{2}  $ must be of the order of the source term   $(1/R)\int_{\mathcal{D}}\mathrm{d}x\mathrm{d}y\ \psi_{1} \partial_{x}\psi_{2}$. Anticipating that typical horizontal scales of the flow structures will be given by $\delta^{1/2} R$, we find that typical variations of the stream function in the lower and the upper layers are related through 
\begin{equation}
\psi_1 \sim  \left(1-\delta \right) \frac{rR}{U} \psi_2 .  
\label{eq:psi1_psi2}
\end{equation} 
We conclude that  $\psi_1 \gg \psi_2$ when $rR /U\gg 1$. At lowest order, only the upper layer is active and the flow can be described by  a 1-1/2 layer quasi-geostrophic model:  
\begin{equation}
\partial_{t}q_{1}+J\left(\Psi_{1},q_{1}\right)=0\ ,\label{eq:dyn_pv_12}
\end{equation} 
with  the notation $\Psi_1=\psi_1-Uy$ and  with
\begin{equation}
q_{1}=\nabla^2 \Psi_{1}-\frac{\Psi_{1}}{\delta R^{2}} .\label{eq:pv12}
\end{equation}

Let us now estimate the typical time scale for the energy evolution. Anticipating the emergence of ribbons, we assume that the total energy is dominated by the potential energy $ E_{tot} \sim L_y^2\Psi_1^2 / R^2$.  This energy should decay with time according to Eq. (\ref{eq:Energy-dynamics-tot}).  We use the scaling   Eq. (\ref{eq:psi1_psi2})  to estimate  $ \int_{\mathcal{D}} \mathrm{d} x  \mathrm{d} y \left(\nabla \psi_2 \right)^2 \sim U^2 \psi_1^2 L_y / (r^2\delta^{1/2} (1-\delta)^2 R^3) $.  Introducing the dissipation time $t_{diss}$ such that $dE_{tot}/dt \sim E_{tot} / t_{diss}$, and assuming $\psi_1\sim \Psi_1$ we get  
\begin{equation}
t_{diss} \sim  \frac{1}{r} \frac{1-\delta}{\delta^{1/2}} \left(\frac{rR}{U}\right)^2\frac{L_y}{R} \label{eq:tfriclarge}
\end{equation}
This leads to a surprising result: in the large bottom friction limit, the typical time scale for the evolution of the quasi-stationary large scale flow is proportional to the bottom friction coefficient. This estimate for the dissipation time Eq. (\ref{eq:tfriclarge})  justify  our choice for  $F_{diss}$ Eq. (\ref{eq:t_diss_F})  in  the limit $rR/U \gg 1$ and $\delta^{1/2} R \ll L_y$. The main caveat of this argument is to assume $\psi_1\sim\Psi_1$ which can not be valid at short time (when the instability grows) and at large time (when the perturbation has almost annihilated the mean flow). However, we will show that this provides a reasonable scaling to interpret the numerical simulations. In addition, same argument applied to the energy budget of the perturbed flow Eq. (\ref{eq:Energy-dynamics-tot}), without assuming $\psi_1\sim \Psi_1$, would show that $t_{diss}$ is the typical time scale for the growth of the potential energy of the perturbed state.%One caveat in the estimate above is  hat the hypothesis $\psi \sim \Psi$ is not valid until the potential vorticity has ben sufficiently stirred( $t\ll t_{diss}$), and it is not valid when the total flow vanish ($t\gg t_{diss}$). 
%A similar scaling argument would show that $t_{diss}$ is  a typical time for the growth of the potential energy for the perturbed state.

\subsection{Cascade phenomenology for quasi-geostrophic models}\label{sec:cascade}

The flow in the  large friction limit $rR/U\gg 1$ and in the low friction limit $rR/U\ll 1$ are both described at lowest order by a one layer  flow model:
\begin{equation}
\partial_t q +J(\psi,q)=0,\quad q=\nabla^2 \psi -\lambda_d^{-2} \psi  \ . \label{eq:generic}
\end{equation}
We recover the barotropic dynamics Eq. (\ref{eq:baro}) when $\lambda_d=+\infty$ and the 1-1/2 layer quasi-geostrophic dynamics Eq.  (\ref{eq:dyn_pv_12}-\ref{eq:pv12}) when $\lambda_d= \delta^{1/2} R$.\\

We consider Eq.  (\ref{eq:generic}) with an arbitrary $\lambda_d$ and we introduce the relative vorticity $\zeta=\nabla^2\psi$. 
At spatial scales much smaller than $\lambda_d$ the potential vorticity $q$  is dominated
by the relative vorticity and the dynamics is  given by 2d Euler equations:
\begin{equation}
\partial_{t}\zeta+J(\psi,\zeta)=0.\label{eq:EulerEq}
\end{equation}
Classical arguments~\citep{Fjortoft53,Kraichnan_Phys_Fluid_1967_2Dturbulence} predict a direct cascade
of enstrophy $Z=\int_{\mathcal{D}}\mathrm{d}x\mathrm{d}y\ \zeta ^{2}/2$ and an inverse cascade of kinetic energy $KE=-\int_{\mathcal{D}}\mathrm{d}x\mathrm{d}y\ \psi\zeta/2$. In the freely evolving case, one expects a decrease of the energy k-centroids $k_E=\int \mathrm{d} k \ k \mathcal{E}(k) / E $  until the energy is condensed at the domain scale, and a concomitant increase of the enstrophy $k$-centroids  $k_Z=\int \mathrm{d} k \ k \mathcal{Z}(k) / E $, where $\mathcal{E}(k)$ and $\mathcal{Z}(k)$ are the energy and enstrophy spectra~\citep{NazarenkoPRL09}.\\ %We note that the barotropic dynamics always fall into that case since $\lambda_d=+\infty$. From now on  we consider the case $\lambda_d=\delta^{1/2} R \ll L_y$.  \\ 
%%ici \cite{ici} stop%
At spatial scales much larger than $ \lambda_d$, the dynamics Eq. (\ref{eq:generic}) is the so-called planetary
geostrophic model~\citep{LarichevMcWilliams91}: 
\begin{equation}
\partial_{\tau}\psi+J(\zeta,\psi)=0,\label{eq:planetary-qg}
\end{equation}
with $\tau=\delta R^2 t$. The role of $\zeta$ and $\psi$ are switched with respect to the Euler dynamics. Same arguments than in the Euler case predict a direct cascade of kinetic energy $KE$, and an inverse cascade of potential energy~\citep{Pierrehumbert94,SmithEtAl} $APE=\int_{\mathcal{D}}\mathrm{d}x\mathrm{d}y\ \psi^2)/(2\lambda_d^2)$. In the freely evolving case, one expects that the available potential energy centroids will go to large scale until condensation at the domain scale. Meanwhile, the kinetic energy centroids should go to small scales.  

We see that  both the small scale  limit described by Eq. (\ref{eq:EulerEq}) and in the large  scale limit described Eq. (\ref{eq:planetary-qg}), the kinetic energy is expected
to pile up at scale $\lambda_d=\delta^{1/2}R$. 

We also note that the concomitant condensation of  potential energy at the domain scale with a direct cascade of kinetic energy (halted around the scale $\delta^{1/2} R$) is necessarily associated with the formation of large regions of homogenized streamfunction at a coarse grained level (or equivalently homogenized potential vorticity). In other words, the streamfunction gradients are expelled at the boundary between regions of homogenized potential vorticity. This justifies  with a dynamical  point of view the emergence of ribbons.  Another complementary point of view is to say that the dynamics tends to homogenize the potential vorticity field, but that a complete homogenization would not be possible due to energy conservation.  In the limit $\delta^{1/2} R\ll L_y$, the dynamics will therefore tend to form at least two regions of  homogenized potential vorticity at the domain scale, which  allows to sustain a large scale available potential vorticity field, while allowing for potential vorticity homogenization almost everywhere. 
 
Typical values of the potential vorticity in the region where it is homogenized can be estimated as $\mathcal{Q}_{1} \sim UL / \delta R^{2}$. We see from Eq. (\ref{eq:generic}) that sufficiently far from the interface, between two regions of homogenized potential vorticity the streamfunction
is also a constant  with $\Psi_{1}\sim \delta R^2 R^{2}\mathcal{Q}_{1}\sim UL_y$.
The interfaces between different regions of homogenized potential vorticity correspond
therefore to jumps of the streamfunction, which occurs at a typical
scale $R$. This corresponds to strong an localized jets with velocity
$V\sim\Psi_{1}/R\sim UL/R$. The length of these jets is of order of the domain size 
$L_y$, much larger than their width, of order $\delta^{1/2} R$, hence the term
{}``ribbons''.  

To conclude,  the flow should self-organize into a large scale structure with velocity variations at the scale of the domain $L_y$  in the low bottom friction limit $rR/U\ll 1 $, and form ribbons of width  $\delta ^{1/2} R$ and length $L_y$ in the large bottom friction limit $rR/U\gg 1$. More detailed predictions for this large scale flow structure can be obtained in the framework of equilibrium statistical mechanics, as discussed in the following subsection. 

\subsection{Statistical mechanics predictions for the large scale flow structure}\label{sec:statmech}

Turbulent dynamics stretches and folds potential vorticity filaments 
which thus cascade towards smaller and smaller scales.
This stirring tends to mix the potential vorticity field at a coarse-grained level, even in the absence of small scale dissipation. If there is no energy constraint and if there is enough stirring, the potential vorticity field should be fully homogenized just as in the case of a passive tracer. By contrast, complete homogenization can not be achieved if there is an energy constraint, which leads to non trivial large scale flow structures, and statistical mechanics gives a prediction for  such large scale flows. The aim of this subsection is to review existing results on the statistical mechanics theory for one layer quasi-geostrophic models that will be useful to interpret our numerical results. 

\subsubsection{Miller-Robert-Sommeria (MRS) theory for a barotropic model}  \label{sub:mecastatonelayer}

The theory was initially developed by \citet{SommeriaRobert:1991_JFM_meca_Stat,milleretal1992}, and will be referred to as the MRS theory in the following. We provide here a short and informal presentation of this approach --- see also reviews by Refs.~\citep{MajdaBook,marston2011,BouchetVenaillePhysRep,lucarinietal2013}.
  
The theory provides a variational problem that allows to compute the most probable outcome of turbulent stirring  at a macroscopic (or coarse-grained) level among all the microscopic configurations of the flow that satisfy the constraints of the dynamics given by the conservation of the energy and of the global distribution of potential vorticity levels. Large deviation theory allows then to show that an overwhelming number of microscopic states corresponds to the most probable macroscopic state. The only assumption is ergodicity, i.e. that there is sufficient mixing in phase space for the system to explore all the possible configurations given the dynamics constraints. 

In the case of a one layer quasi-geostrophic flow described by Eq. (\ref{eq:generic}), the input of the theory is given by the energy of the flow  $E$ and the initial fine-grained (or microscopic) potential vorticity distribution $\gamma(\sigma)$. The output of the theory is a field  $p(x,y,\sigma)$ that gives the probability density function to measure
a potential vorticity  level $\sigma\in\Sigma$ in the vicinity of the point $(x,y)$. This field defines a  macroscopic state of the system, which allows to keep track of the dynamical constraints. The computation of the equilibrium state amounts to find the field
$p$ that maximizes a mixing entropy $\mathcal{S}=-\int_{\Sigma}\mathrm{d}\sigma\int_{\mathcal{D}}\mathrm{d}x\mathrm{d}y\ p\ln p$
with the constraints given by dynamical invariants expressed in term of $p$. This entropy counts the number of micro states associates with a given macro state $p$~\citep{SommeriaRobert:1991_JFM_meca_Stat,milleretal1992}. 
The constraints are given by the conservation of the global distribution  $\gamma(\sigma)=d_{\sigma}$ with $ d_{\sigma}[p]=\int_{\mathcal{D}}\mathrm{d}x\mathrm{d}y\int \mathrm{d}\sigma p$, and  the energy conservation $E=\mathcal{E}[p]$ with $\mathcal{E}[p]=- \int_{\mathcal{D}}\mathrm{d}x\mathrm{d}y\int \mathrm{d}\sigma \sigma  p \psi $. Note that  the energy constraint is obtained by assuming that the energy of local vorticity fluctuations is negligible. The validity of this  mean-field treatment can be proven using large deviation theory.  The potential vorticity field of the equilibrium state is $\overline{q}=\int_{\Sigma}\mathrm{d}\sigma\ \sigma p$, and the streamfunction is obtained by inverting  $\overline{q}=\nabla^2\psi-\lambda_d^{-2} \psi$.
We stress that the theory applies for flows without small scale dissipation. In the presence of small scale dissipation, the predictions of the theory is expected to be valid  only if the typical time scale for self organization of the flow is much smaller than the typical time associated with small scale dissipation. We also note that in that case, once the flow is self-organized, small scale dissipation smears out local fluctuations of the potential vorticity field so that the microscopic potential vorticity field $q$ actually tends to the macroscopic field  $\overline{q}$.

The equilibrium state is always characterized by a monotonous functional relation $\overline{q}=g(\psi)$~\citep{SommeriaRobert:1991_JFM_meca_Stat,milleretal1992}. This function $g$ depends only on the dynamical invariants. At this stage two approaches could be followed. A first approach is to consider  $E$  and $g(\sigma)$ as given, to compute the function $g$, and the flow structure associated with the corresponding equilibrium state. A second approach is to assume a given $q-\psi$ relation, and to compute the MRS statistical equilibria associated with this relation. This second approach has made possible several analytical  results in the last decade, and we will rely on these results to interpret our simulations.  
 
 Although computation of the equilibrium state is a difficult task in general, several analytical results can be obtained in limit cases~\citep{BouchetVenaillePhysRep} for a detailed discussion. For instance, whatever the initial distribution of potential vorticity levels, it can be shown that  low energy state are always characterized at lowest order by a linear $q-\psi$ relationship, whose coefficient only depend on the total energy, the total enstrophy and the circulation~\citep{BouchetVenaillePhysRep}.  Here low energy means that the energy of the flow is much smaller than the maximum admissible energy for a given potential vorticity distribution. In our case the initial total energy is of the order of $U^2L_y^3L_x/ R^2$. It is not difficult to construct a state, with the same global  distribution of potential vorticity levels, that is characterized by an energy that scales as $U^2L_y^3L_x/(\delta R^4)$, which is therefore much larger than the initial energy provided that $\delta R \ll L_y$%\footnote{To construct such a state one just needs to consider (\ref{eq:Energy}). This gives an estimate of the energy of a flow obtained by rearranging the initial potential vorticity field in such a way that its typical variations are given by $ \mathcal{Q}_1$, assuming $ \mathcal{Q}_1\sim UL_y/(\delta R^2)$, i.e. that the potential vorticity has not been ``mixed", but only rearranged at large scale}.
This justifies the  low energy limit for the weak friction case.
   
Such a low energy limit allows us to compute analytically phase diagrams for the flow      
structure and to describe how this flow structure changes when the energy or the enstrophy of the flow are varied. For instance statistical equilibria associated with a linear $q-\psi$ relation have been classified for various flow model in an arbitrary close domain~\citep{chavanis1996,VenailleBouchetJSP} and on a channel~\citep{corvellec2012}.  In particular,  it was shown in these studies that when the flow domain is sufficiently stretched in the $x$ direction, then the equilibrium state is a dipolar flow. 
% to interpret the numerical results in the low friction case $rR/U \ll 1$.\\   
 
 \subsubsection{Application to the 1-1/2 layer quasi-geostrophic model} \label{sub:mecastatonelayer1/2}
 
 In the large friction limit $rR/U\gg 1$, our justification for the relevance of the ``low energy limit" of the previous subsection is no more valid, since this justification relied on the estimate Eq. (\ref{eq:Energy}) with the underlying assumption that the flow is fully barotropic. We have shown previously that in the large friction limit, the flow is not barotorpic, but is described at lowest order by the  1-1/2 quasi-geostrophic dynamics Eq. (\ref{eq:generic}) with  $\lambda_d=\delta^{1/2} R$. When  $\delta^{1/2} R \ll L_y$,  i.e.  when the Rossby radius of the upper layer is much smaller than the domain scale, it has been shown by \citet{BouchetSommeriaJFM02} that a  class of equilibrium state different that the low energy states of the previous section  can be computed analytically.  Assuming that the  $q-\psi$ relation is $\tanh$-like, they showed that the 
equilibrium state is composed of two subdomains with homogenized potential
vorticity separated by jets of width $\delta^{1/2} R$ at their interface, see
also~\citep{WeichmanPRE06,VenailleBouchetJPO10}.  Statistical mechanics also predict in that case that the interface between the two regions of homogenized potential vorticity should be minimal, just as bubbles in usual thermodynamics.   A key assumption for these results is that the $q-\psi$ relation of the equilibriums state has a tanh-like shape.  In the case of an initial distribution $\gamma(\sigma)$ with only
two levels of potential vorticity, it can be shown than the $q-\psi$ relation is given exactly by a tanh function~\citep{BouchetSommeriaJFM02}. \citet{BouchetSommeriaJFM02} conjectured that there exist a much larger  class of  initial energy $E$  and of fine-grained potential vorticity distributions $\gamma(\sigma)$ that leads to a tanh-like shape  for the $q-\psi$ relation  at equilibrium. Our phenomenological arguments above and our numerical results below suggest that the dynamics is indeed attracted toward  a quasi-stationary state characterized by such a tanh-like relation in a case where the initial distribution of potential vorticity levels is far from a double delta function, see Fig. \ref{fig:pv_init}. 

\section{Numerical results}
\label{sec:test}

\subsection{Numerical settings}

Quasi-geostrophic simulations are performed using the same numerical model as in \citet{nadeaustraub2009}.  No normal flow and slip conditions are imposed at lateral walls. We use third order Adams-Bashforth scheme for time derivatives, center differencing in space, \citet{arakawa1966} for the Jacobian, and a multigrid method for the elliptic inversions.  Momentum conservation is achieved following a procedure similar to that of \citet{mcwilliamshollandetal1978}, using the zonal momentum equation integrated over a latitude circle in the channel.

 To trigger the instability, we considered an initial streamfunction perturbation
given by a random velocity field with random phases and a gaussian
spectrum of width $\Delta k=2$ and peaked at $k=6$; the perturbation
where such that $\Psi_{1}^{init}k\ll U$. As we will see in the high bottom friction limit, the dynamics required thousands 
of eddy turn-overtime, hence the moderate horizontal resolution. The initial condition for the potential vorticity fields $q_1,q_2$ is represented on Fig. \ref{fig:pv_init}.

\begin{table}
\begin{center}
\begin{tabular}{|l c c|}
\hline
Parameter & \multicolumn{2}{c|}{ $ {\rm Value} $ }\\
\hline \hline
& \\

Imposed velocity  & \multicolumn{2}{c|}{$ U  = 1 {\rm m.s}^{-1} $}\\

Channel width & \multicolumn{2}{c|}{$ L_y = 900 {\rm km}$}\\

Fractional depth of the upper layer & \multicolumn{2}{c|}{$ \delta  = 0.2  $}\\

Rosby radius  & \multicolumn{2}{c|}{$R/L_y = 0.1 $} \\

Channel  aspect ratio & \multicolumn{2}{c|}{$L_x/L_y = 5/3$} \\

Bottom friction coefficient & \multicolumn{2}{c|}{$r U/R$ from $0$ to $40$} \\

Horizontal resolution & \multicolumn{2}{c|}{$\Delta x = \Delta y = 1.7 {\rm km} $} \\

Bi harmonic dissipation coefficient & \multicolumn{2}{c|}{$A_h= 1. 10^8 {\rm s}^{-1} {\rm m}^4$}\\ [2ex]

\hline
\end{tabular}
\end{center}

\caption{Model parameters for the reference simulations. Other simulations have been performed by varying $R/L_y$ and $Lx/L_y$.}
\label{table:parameters1}
\end{table}

There are five adimensionalized parameters in this problem:  the adimensisonalized bottom friction  coefficient $rR/U$, the aspect ratio $L_x/L_y$, the adimensionalized internal Rossby radius of deformation $R/L_y$, the ratio  $\delta$ of the upper layer depth with the total depth, and  the  Reynolds number based on the small scale dissipation coefficient $A_h$. The
small scale dissipation coefficient is adjusted to the lowest necessary value to ensure
convergence of the simulation for a given resolution.  \citet{ArbicRibbon} did show that the
result of such simulation does not depend strongly (at least qualitatively)
on the form chosen for the small scale dissipation term.  We also checked that our results were not dependent on the chosen resolution. Consistently with the exponential stratification observed in most parts of the oceans, we consider that the upper layer is thin compared to the lower layer, with  $\delta = 0.2$, and this parameter will be constant for all the simulations. This choice is also reasonable to test the scaling predictions obtained for $\delta\ll 1$.  There remains three parameters. The main control parameter is $rR/U$ which is varied from $0$ to $40$, in order to test our scaling predictions obtained for $rR/U\ll 1$ and $rR/U\gg 1$.   We considered a ratio $R/L_y=0.1$ for the reference case (which corresponds to $\delta^{1/2} R/L_y=0.004$) but also looked at the effect of decreasing  this parameter. In any case this parameter can be considered to be much smaller than one.  We finally considered aspect ratio $L_x/L_y=5/3$ for the reference case, which corresponds to a grid $897X513$ in physical space. We explored the effect of varying the domain aspect ratio, but always in the regime $L_y\sim L_x$. These parameters are summarized in table \ref{table:parameters1}. 

% [Not sure this paragraph is needed.] We discuss in subsection \ref{sub:energynum}  how the vertical structure of the flow and the typical time scale for the energy variations are affected by the parameter $rR/U$. We then address in subsection \ref{sub:weakfrictionnum}  the main qualitative properties of the observed flow  when the parameter $rR/U$ is small or of order one. We finally describe in more details the ``ribbon" regime, when $rR/U$ is large, in subsection \ref{sub:ribbonnum}. 
  
\subsection{The role of bottom friction} \label{sub:energynum}

\begin{figure}
\begin{center}
\includegraphics[width=\textwidth]{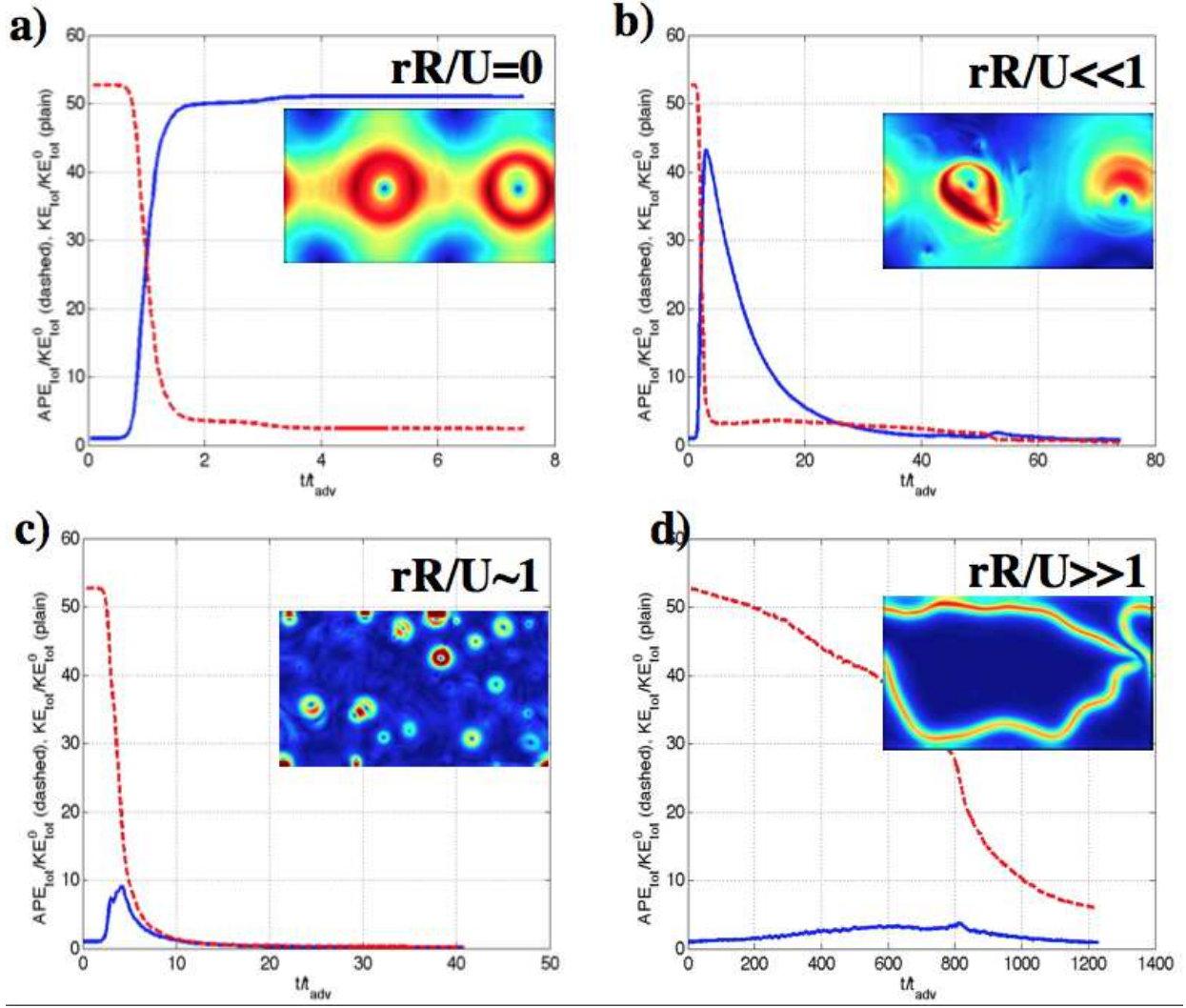}
\end{center}
\caption{ a) Temporal evolution of the total kinetic energy $KE_{tot}$ and of the total potential energy $APE_{tot}$ in the case $rR/U$. Time unit is normalized by $t_{adv}=L_y/U$. The field in inset represents a snapshot of the velocity  modulus during the kinetic energy decay.  b) idem $rR/U=0.004$ c) idem for $rR/U=0.5$ d) idem for $rR/U=40$. Note that the flow structures in each regime are similar to Fig. 7 of  \citet{arbicflierl2004}. \label{fig:timeserie}}
\end{figure}

\begin{figure}
\begin{center}
\includegraphics[width=0.7\textwidth]{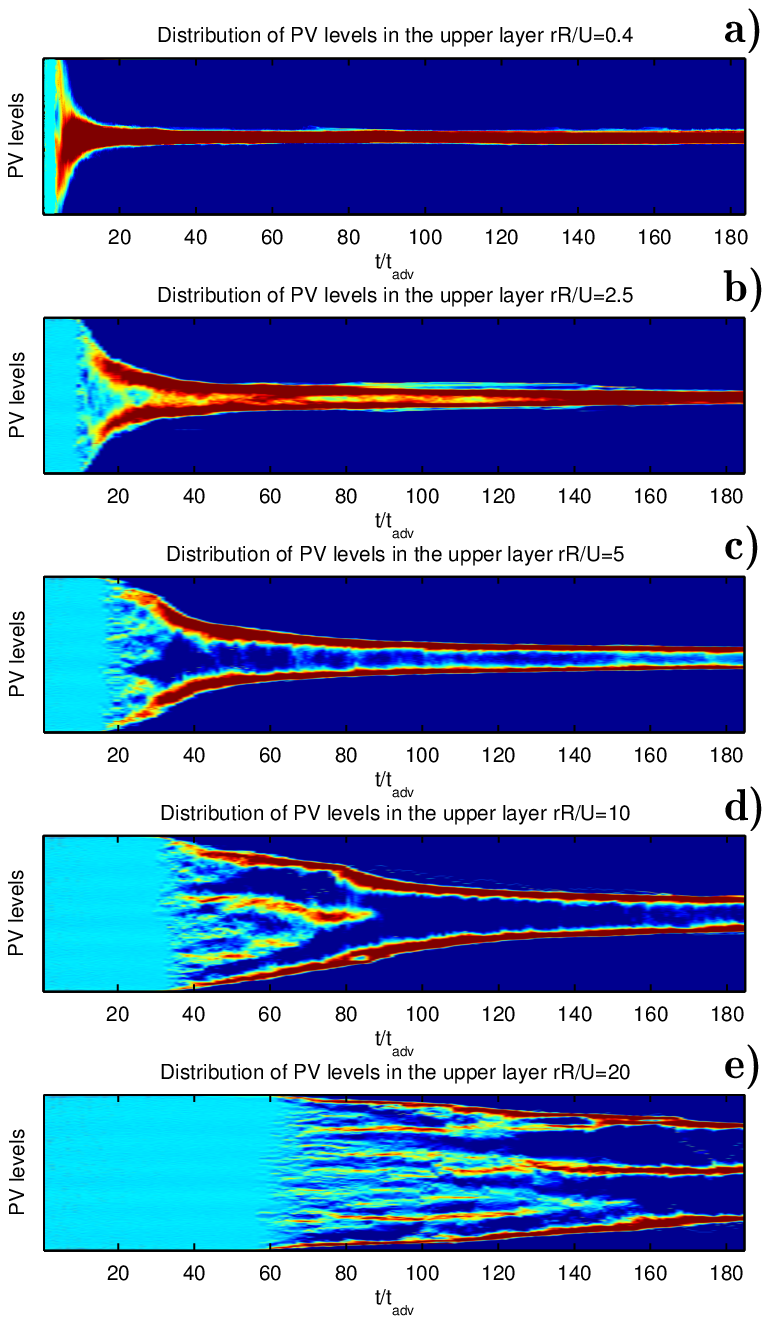}
\end{center}
\caption{ Routes towards potential vorticity homogenization depending on bottom friction. Each panel represents the temporal evolution of the global distribution of potential vorticity levels in the upper layer.  Time is adimensionalized by $t_{adv}=U/L_y$ \label{fig:homog}.}
\end{figure}

\subsubsection{Energy decay and potential vorticity homogenization}

 We first discuss reference simulations for which  the  aspect ratio is $L_x/L_y=5/3$ and the Rossby radius is $R/L_y=0.1$.  We present Fig. \ref{fig:timeserie} the temporal evolution of the total kinetic energy  $KE_{tot}=KE_{1tot}+KE_{2tot}$  and of the total available potential energy  $APE_{tot}$ defined Eq. (\ref{eq:Energy_tot}), for various values of the  bottom friction coefficient $rR/U$.  We see that in any case, the total available potential energy $APE_{tot}$ decreases and eventually vanish. We distinguish three regime for the temporal evolution of the kinetic energy   $KE_{tot}$  
 \begin{enumerate}
 \item the initial growth of  $KE_{tot}$
 \item the saturation  regime  where $KE_{tot}$ reaches its maximal value
 \item the decay of  $KE_{tot}$ due to bottom friction (except when when $rR/U = 0$).
 \end{enumerate}
 
 As explained before, the decay of the total energy to zero indicates the potential vorticity field is fully homogenized, so that the perturbation has cancelled the effect of the imposed mean flow. Remarkably, the different routes towards complete homogenization and the time scales associated with it are completely different depending on the value of $rR/U$, which appears clearly on the temporal evolution of the global distribution of potential vorticity levels in the upper layer, see Fig. \ref{fig:homog}.  The observed flow structures during this energy decay also strongly depends on the coefficients $rR/U$ as shown on the insets of Fig. \ref{fig:timeserie}. In the weak friction case, the flow is a large scale dipolar vortex condensed at the domain scale. In the large bottom friction limit the flow is a ribbon of kinetic energy of width given by $\delta^{1/2} R$, and in the intermediate bottom friction limit the flow is  made of isolated vortex whose size is of the order of the Rossby radius of deformation $R$. We note that all these flow configurations are qualitatively similar to the one reported in the doubly periodic case by \citet{arbicflierl2004}.

 \subsubsection{Estimate for the dissipation time}
 
\begin{figure}
\begin{center}
\includegraphics[width=\textwidth]{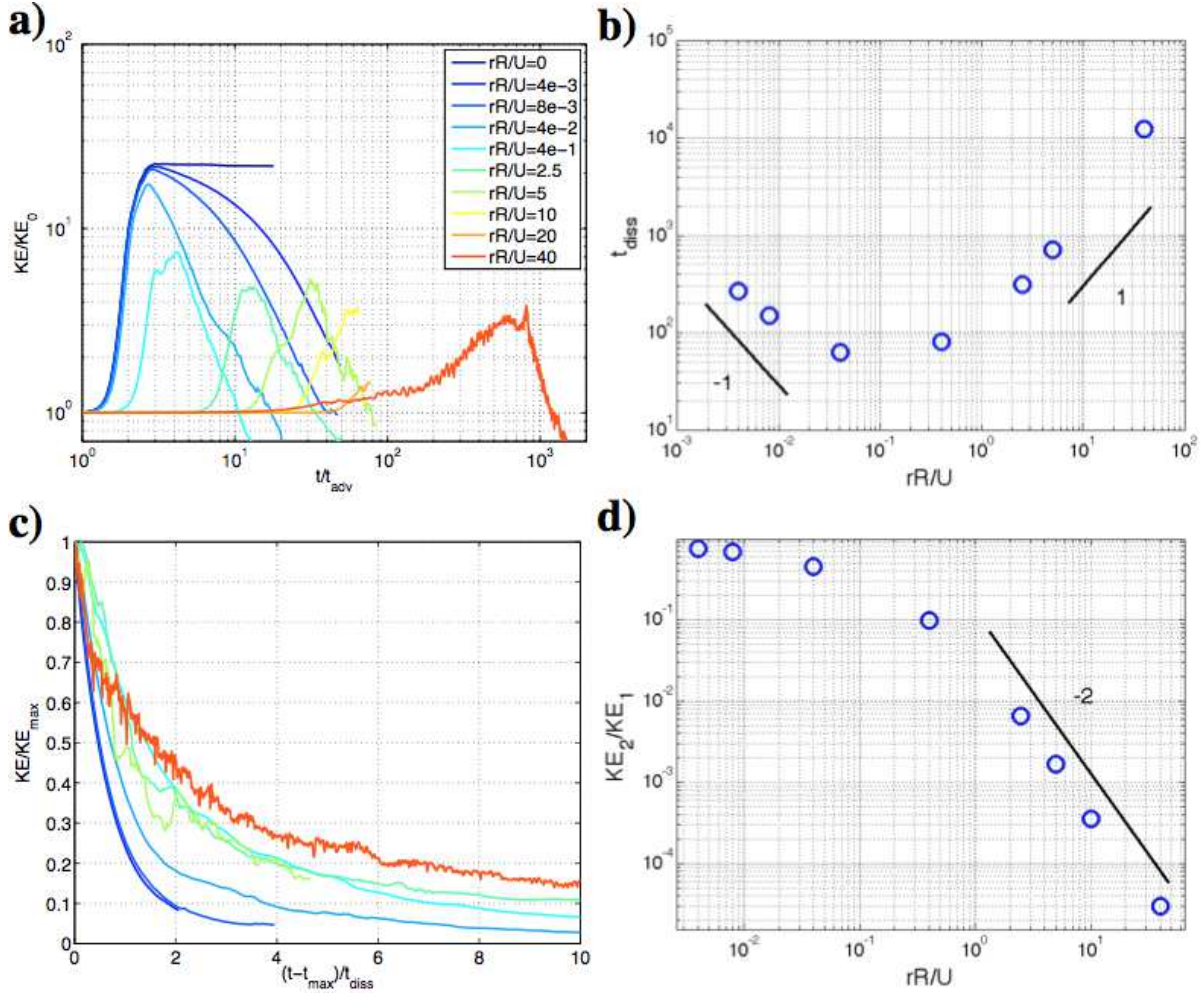}
\end{center}
\caption{ a) Temporal evolution of the total kinetic energy $KE=KE_1+KE_2$. The kinetic energy is normalized by its initial value, and the time scale is normalized by the advection time $t_{adv}=L_y/U$. The logarithm scale is used in order to see all the runs on the same plot.  b) Estimation of the dissipation time in the numerical experiment (see text for details).  c)  Temporal decay of the kinetic energy $KE_{tot}$. The time series are the same than on panel a, but the $KE_{tot}$ is normalized for each run by its maximum value, time coordinate is normalized by the dissipation time defined Eq. (\ref{eq:t_diss_general}-\ref{eq:t_diss_F}), and the time origin has been translated for each run so that $t=0$ corresponds to the time where the kinetic energy is maximal.
 \label{fig:friction}}
\end{figure}

We compare on  Fig. \ref{fig:friction}-a  the temporal evolution of the total kinetic energy $KE_{tot}$ for various values of $rR/U$.  Clearly, the time scales for this temporal evolution strongly depend on the value of $rR/U$.  Let us first discuss the initial energy growth. It is a classical result that in the weak friction regime  $rR/U\ll 1$  the typical time for baroclinic instability scales as $R/U$, hence  the  initial collapse of all the curves that belong to this regime on Fig. \ref{fig:friction}-a. For the same reason, the saturation of the instability due to self-organization following turbulent stirring always occurs at a time scale of the order of the advection time $t_{adv}=L_y/U$ in this low friction regime.  By contrast, in the high friction limit $rR/U\gg 1$, a direct computation of the linear baroclinic instability would show that this instability increases linearly with the bottom friction coefficient $r$. In addition, our estimate for the non-linear growth of the energy of the perturbation (see  the end of subsection \ref{sub:surface_intensification}) lead to a time $t_{diss}$ that also scales linearly with the bottom friction coefficient $r$. These predictions agrees qualitatively with the fact that  kinetic energy peaks occurs at larger time with increasing bottom friction coefficient $r$ on Fig.  \ref{fig:friction}-a.

We focus now on the kinetic  energy decay. For a given value of the parameter $rR/U$, we estimate  on  Fig.  \ref{fig:friction}-b the decay time $t_{diss}$ as the time interval between the kinetic energy maximum $KE_{max}$ and $KE_{max}/4$.  We see that the predictions for this dissipation time given by  Eq. (\ref{eq:t_diss_general}-\ref{eq:t_diss_F}) yields a good qualitative understanding of the numerical simulations in the low bottom friction regime ($t_{diss}\sim 1/r$) and the large bottom friction regime ($t_{diss}\sim r$). 
In order to test in more details these predictions for the energy dissipation time scale, we plot on Fig. \ref{fig:friction}-c the temporal evolution of the kinetic energy starting from $t_{max}$, the time when the maximum total kinetic energy has been reached, by renormalizing time unit with the dissipation time $t_{diss}$ given by   (\ref{eq:t_diss_general}-\ref{eq:t_diss_F}), for each value of the parameter $rR/U$.  Remarkably, and despite the four decades range for $rU/R$, all the curves for the energy decay collapse qualitatively well. This good collapse confirms  that not only the scaling obtained in the limit cases are correct, but prefactors are also qualitatively correct.

 \subsubsection{Vertical flow structure}
  
 We show on Fig. \ref{fig:friction}-d the ratio $\delta KE_{2tot}/ (1-\delta )KE_{1tot}$ of the total kinetic energy in each  layer normalized by the depth of these layers, as a function of the parameter $rR/U$.  We expect from subsection \ref{sub:barotropization} that this energy ratio  tends to one when $rR/U\ll 1$, i.e. that the flow has become barotropic.  We expect from the scaling Eq. (\ref{eq:psi1_psi2}) that this energy ratio should scale as  $ \sim (rR/U)^{-2}$ for large $rR/U$. We see a very good agreement between these predictions and  our our numerical results on Fig. \ref{fig:friction}-d. We stress  that both scalings  are based on the fact that the flow is self-organized into a quasi-stationary states. This contrasts with the scaling  $\delta KE_{2tot}/ (1-\delta )KE_{1tot} \sim (rR/U)^{-4/3}$ proposed by \citet{arbicflierl2004} by revisiting a  cascade argument by \citet{heldlarichev1996}. We believe that their scaling is relevant to describe the vertical structure of the flow for $rR/U\gg1$ provided that the potential energy length scale remains smaller than the domain size. Since the potential energy length scale increases with $rR/U$, this scaling should break at some point.  In any case, both our scaling and the scaling of \citet{arbicflierl2004}  predict that  the dynamics is well described by a 1-1/2 quasi-geostrophic model in  the limit of large frictions $rR/U \gg 1$, and by a barotropic flow model in the low friction limit $rR/U\ll 1$.  The next two subsections are devoted to the description of the flow structure in both regimes.   
   
 \subsection{Weak friction limit}\label{sub:weakfrictionnum}

\begin{figure}
\begin{center}
\includegraphics[width=\textwidth]{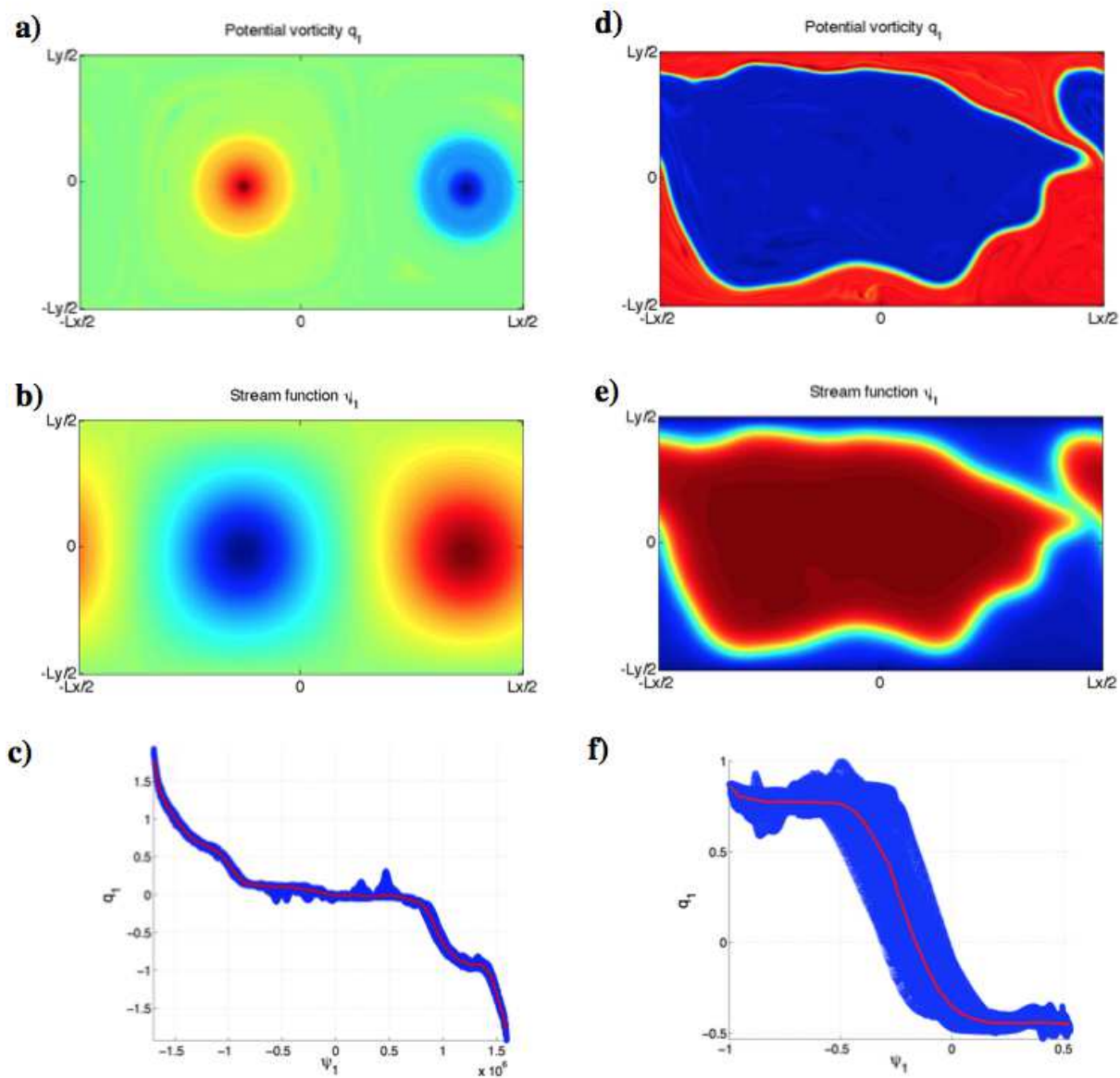}
\end{center}
\caption{ a) Potential vorticity field  in the upper layer  for $rR/U=0$. b) corresponding streamfunction field in the upper layer c) scatterplot of the $q-\psi $ relation associated with a and b. d,e,f) same plots in the case with large bottom friction $rR/U\ll 1$ \label{fig:q_psi}.}
\end{figure}

 We see Fig.  \ref{fig:timeserie}-a that the flow reaches a stationary state  when   $rR/U=0$. We checked that in this state, $80 \%$ of the kinetic energy was in the barotropic mode, which is in agreement with the fact that  barotropization is expected  with corrections of order $\delta$ or $R/L_y$ when $rR/U\ll 1$ and $\delta \ll 1$, see the discussion subsection \ref{sub:barotropization}.  We also note that the initial potential energy reservoir of the baroclinically unstable mean flow ($APE_{tot}^0\gg KE_{tot}^0$) has been  transferred almost totally into kinetic energy, due to the conservation of the total energy $E_{tot}=APE_{tot}+KE_{tot}$.    We see Fig. \ref{fig:q_psi}-a,b  that the corresponding large scale streamfunction and potential vorticity fields are self-organized into a dipolar structure at the domain scale. This dipole is characterized by a monotonous relation between potential vorticity and streamfunction. This functional relation has roughly a $\sinh$ shape. This $\sinh$ shape is different than the linear $q-\psi$ relation that one would expect in a low energy limit for a  initial prescribed potential vorticity distribution. We explained subsection \ref{sub:mecastatonelayer} that the total energy in the  numerical experiment is much smaller than the maximal admissible energy with the same initial global distribution of potential vorticity levels. The reason why a linear $q-\psi$ relation  is not observed here is that the core of the remaining vortices have not been stirred during the turbulent evolution of the flow. This shows a lack of ergodicity for the dynamics, which has  been discussed for instance by \citet{schecter1999}. However, we note that the observed dipolar structure is the flow that would be predicted by the MRS theory applied to the barotropic model in a channel sufficiently stretched in the $x$-direction, as explained in subsection \ref{sub:mecastatonelayer}. 
  
   In the presence of a  weak bottom friction ($rR/U\ll 1$)  the large scale state becomes quasi-stationary and the total kinetic energy decreases with a time scale of the order of $1/r$ until the total energy vanish. By quasi-stationary me mean that there still exist a well defined $q-\psi$ relation, but with superimposed small fluctuations that increase when bottom friction increases.  The total energy decay goes with the homogenization of the potential vorticity fields. This route towards potential vorticity homogenization is illustrated Fig. \ref{fig:homog}-a. We see on this figure the rapid  emergence of one broad central peak for the global potential vorticity distribution, which indicates that the background potential vorticity field is well mixed over a time $t_{adv}\sim L_y/U$, and the width of the peak  decreases more slowly, over a time scale of the order of $1/r$. we also remark that two isolated peaks with large potential vorticity value persists until $ t_{diss}\sim 1/r$. These peaks correspond to the unmixed core of the dipolar structure. The increase of their strength is an artifact due to the use of a biharmonic dissipation operator. This would not occur with viscous dissipation. 
      
 We note that this route towards complete potential vorticity homogenization and dissipation of the energy of the initial baroclinically unstable mean flow is very much like the classical scenario for two-layer baroclinic turbulence: the instability leads to an inverse energy cascade on the horizontal, with barotropization on the vertical, and then bottom friction dissipates the energy of the large scale flow~\citep{rhines1979,Salmon_1998_Book}. % W
      
 When the bottom friction is further increased, the inverse energy cascade is arrested before the flow self-organizes at the domain scale, and the number of vortices increases.  When $rR/U$ is of order one,  the bottom friction time scale $\sim1/r$  is of the order of the linear baroclinic instability time scale $R/U$. One expects therefore that flow structures can not grow larger than the scale of injection, which is the scale of the most unstable mode for linear instability and secondary instabilities, of order $R$.  This explains  the formation of coherent structures of size $R$ on Fig.  \ref{fig:timeserie}-c. These eddies rapidly mix the background potential vorticity field, on the advection time scale $t_{adv}=L_y/U$,  as seen on Fig. \ref{fig:homog}-b.  This is a strongly out-of-equilibrium regime, that can not be described by MRS equilibria. In the doubly periodic case, this regime of vortex kinetics can be statistically steady, and has been studied in detail by \citet{ThompsonYoung06JPO}. In the case of the channel the number of isolated vortices decreases with time until the potential vorticity field is fully homogenized. 
     
\subsection{Large friction limit}\label{sub:ribbonnum}

\begin{figure}
\begin{center}
\includegraphics[width=0.8\textwidth]{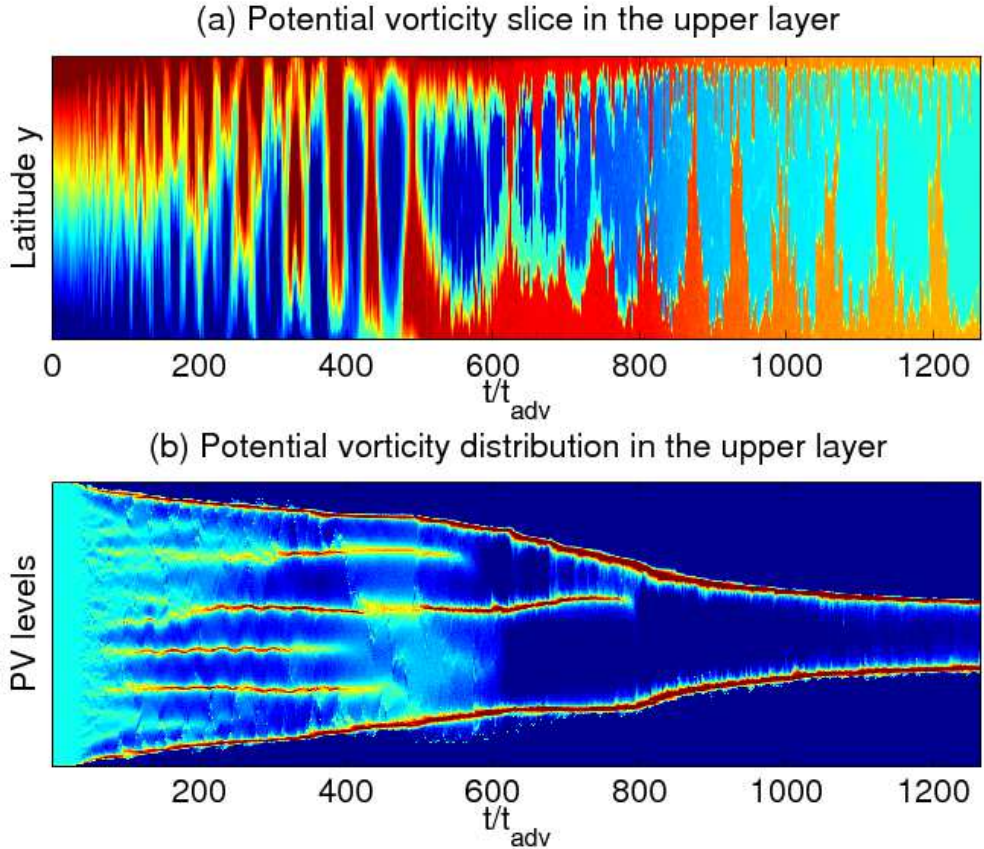}
\end{center}
\caption{ a) Hovm\"oller diagram of a potential vorticity line q(y,t) for a given longitude $x$. b) Temporal evolution of the global distribution of potential vorticity levels. Time is adimensionalized by $t_{adv}=U/L_y$  in both cases.  \label{fig:historibbon}.}
\end{figure}

\subsubsection{Emergence of the ribbons}

\begin{figure}
\begin{center}
\includegraphics[width=\textwidth]{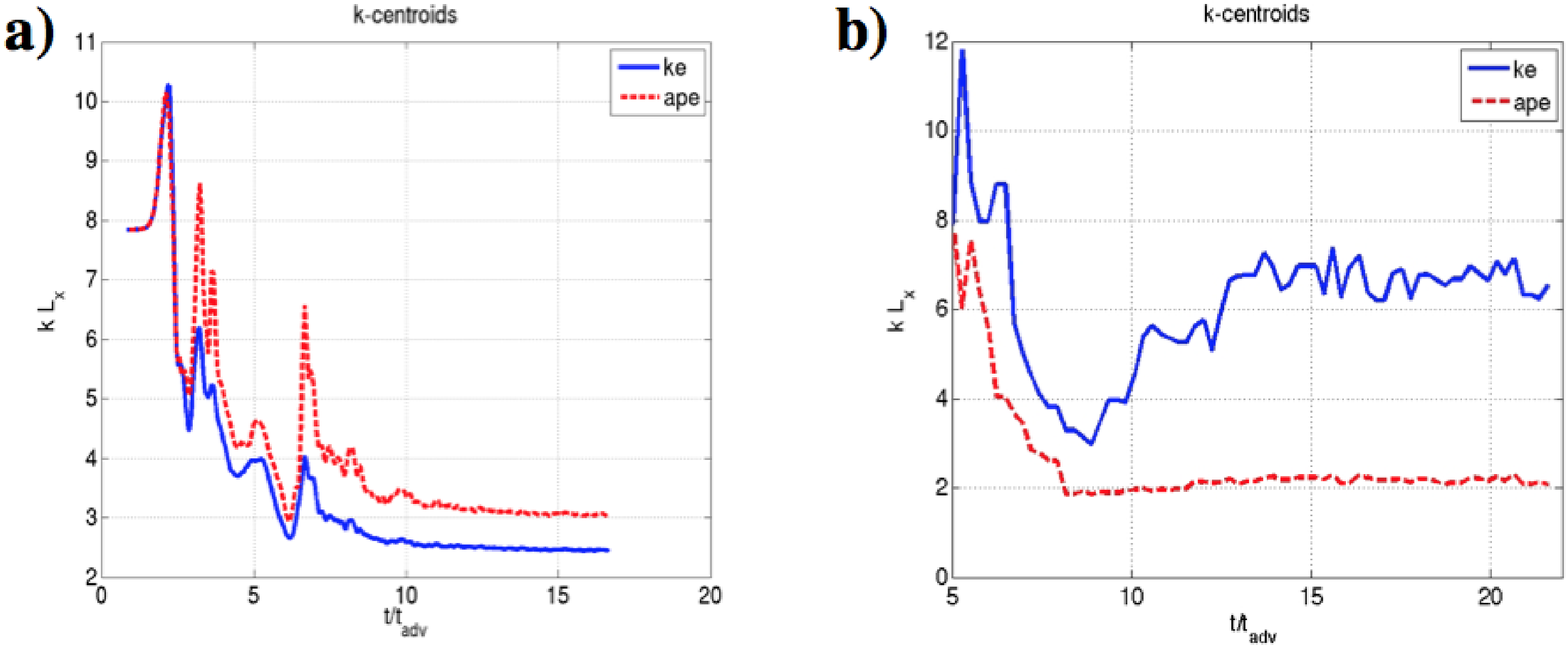}
\end{center}
\caption{a) Initial temporal evolution of the barotropic kinetic energy k-centroid $k_{KE}=\int dk \ k \widetilde{KE}_t (k)/ \int dk \ \widetilde{KE}_t(k)  $ and of the available potential energy centroid  $k_E=\int dk \ k \widetilde{APE}(k)/ \int dk \ \widetilde{APE}(k) $ in the case $rR/U=0$. b) Initial  temporal evolution of the  kinetic energy k-centroid of the upper layer and of the available potential energy k-centroid in the case $rR/U=40$.  Time is adimensionalized in both case by  $t_{adv} =L_y/U$. \label{fig:centroids}}
\end{figure}

A typical snapshot of the potential vorticity field when a quasi-stationary state is reached  is  presented  Fig.  \ref{fig:q_psi}-d for the case $rR/U=40$. Clearly, at sufficiently large time, the flow has reached a state characterized by two regions of homogenized potential vorticity separated by a sharp interface. By sharp we mean that the interface between the homogenized regions is much smaller than the Rossby radius of deformation of the upper layer  $\delta^{1/2} R$. This sharp interface in the potential vorticity field  induces typical variations of streamfunction  at scale $\delta^{1/2}R$ in the transverse direction, see Fig.  \ref{fig:q_psi}-e. The scatterplot of the potential vorticity field and streamfunction field is  plotted on Fig. \ref{fig:q_psi}-f, and shows a tanh-like shape for the $q_1-\psi_1$ relation. The red line is the averaged potential vorticity along one streamline. The presence of fluctuations around this red line indicates that contrary to the case $rR/U=0$, the large scale flow is not exactly a stationary states: the interface meanders intermittently break, and the blobs of potential vorticity exchanged during these breaking events are then stretched and folded in each region of homogenized potential voracity, hence the presence of potential vorticity fluctuations. 

It is notable that the dynamics drives the system towards a state characterized by a `tanh' relation between vorticity and streamfunction, given that the initial potential vorticity field  in the upper layer is a gradient in the meridional direction  presenting no region of homogenized potential vorticity. %To our knowledge, previous numerical works reporting regions of homogenized potential vorticity fields separated by strong jet at their interfaces did consider an initial condition presenting already a few regions of homogenized potential vorticity~\citep{marcus1988}, or only two levels of potential vorticity \citep{bouchet2003}. 
 In that respect, our results support for the the claim of \cite{BouchetSommeriaJFM02} that phase separation of the potential vorticity field into two homogenized regions is a generic feature of 1-1/2 layer quasi-geostrophic equilibria, that does not depend on the particular initial condition when $\delta^{1/2}R/L_y\ll 1$.  

The spontaneous emergence of ribbons  also support the  argument of subsection \ref{sec:cascade} based on cascade phenomenology and  on  potential vorticity homogenization theory.  Indeed, we see Fig. \ref{fig:centroids} a comparison between the temporal evolution of the kinetic and potential energy centroids both in the case $rR/U=0$ (barotropic dynamics at lowest order) and $rR/U=40$ (1-1/2 layer quasi-geostrophic dynamics at lowest order).   In the  case with vanishing bottom friction, the kinetic energy centroid goes to the domain scale and remains there, as expected from inverse energy cascade arguments. In the  case with high bottom friction, the centroids of  potential energy initially goes to large scale, and so does the centroids of kinetic energy (slaved to the inverse cascade of potential energy). But once these centroids have reached the domain scale,  the kinetic energy centroids goes back to smaller scale until a plateau is reached, while the potential energy centroids remains to large scale. This clearly indicates that  streamlines are ``pinched", or expelled at the boundary between regions of homogenized potential vorticity. It was shown by \citet{dritschelscott2011} that such jet sharpening mechanism through turbulent stirring is enhanced by the presence of coherent vortices in the vicinity of the jets. We actually observed the presence of such vortices for values of bottom friction $rR/U $ large but of order one, but these vortices disappeared at large time for $rR/U>10$.

The emergence of the ribbons as a potential homogenization process is conveniently described by a Hovm\"oller diagram  of  Fig. \ref{fig:historibbon}-a  showing the temporal evolution of meridional slice of the potential vorticity profile  $q_1(y,t)$, and by the temporal evolution of the global distribution of potential vorticity levels  shown Fig. \ref{fig:historibbon}-b. Clearly, the dynamics initially form multiples regions of homogenized potential vorticity with ribbons  at their interface, and these regions eventually merge together until two regions of homogenized potential vorticity are formed.

\subsubsection{Ribbon dynamics}

We explained in subsection \ref{sub:mecastatonelayer1/2} that statistical mechanics theory of the 1-1/2 layer model with small $R/L_y$ predicts not only the ultimate formation of two regions of homogenized potential vorticity, but also the organization of these regions into a configuration that minimizes the length of their interface.  Clearly, the interface perimeter of the potential vorticity  field in Fig. \ref{fig:q_psi}-d is not minimal. Moreover, a movie would reveal that this interface is permanently meandering, and sometimes even breaks locally. Indeed, the jets at the interface between the regions of homogenized potential vorticity field are characterized by a strong vertical shear, and are therefore  expected to be baroclinically unstable. This instability is actually a mixed  barotropic-baroclinic instability, since the jets have an horizontal structure.  To check that the meanders were due to the existence of a vertical shear,  we ran a numerical simulation of the 1-1/2 quasi-geostrophic dynamics taking the potential vorticity field of Fig. \ref{fig:q_psi}-d as an initial condition. This amounts to impose $\psi_2=0$, and therefore precludes any baroclinic instability. In those freely evolving simulations  the interface did stop meandering and the flow did reach a stationary state. We also observed that the interface was eventually smoothed out in the freely evolving 1-1/2 layer  quasi-geostrophic simulations, while the interface remains sharp throughout the flow evolution when baroclinic instability is allowed, as seen on the Hovm\"oller diagram \ref{fig:historibbon}-a. We conclude that in the limit of large  bottom friction, there is a competition between baroclinic instability that tends to increase the interface perimeter between regions of homogenized potential vorticity, and the  dynamics of the inviscid 1-1/2 layer quasi-geostrophic dynamics that tends to minimize this interface.

Baroclinic instability of the ribbons is the mechanism that allows to reduce little by little the potential vorticity jumps across the ribbons, at a time scale given by $t_{diss}\sim r RL_y/(\delta^{1/2}U^2)$. This time scale is of the order of  the slow variations of the potential vorticity interface at large time in  the Hovm\"oller diagram \ref{fig:historibbon}-a.  We see on Fig. \ref{fig:historibbon}  that once two regions of homogenized potential vorticity are formed, the value of the potential vorticity jump $\mathcal{Q}_{1jump}$  between  the homogenized regions decreases exponentially, with an e-folding depth of the order of the decay time for the kinetic energy $t_{diss}$. The corresponding flow structure (i.e. meandering jets with a ribbon shape)  remains the same, but the strength of the jet also decreases in time, since $U_{jet} \sim \delta^{1/2} R \mathcal{Q}_{1jump}$.\\
  %We also note that the slow decrease of the potential vorticity jump between both regions is associated with the slow decrease of the total potential energy  Fig. \ref{fig:friction}-c.

\subsubsection{A competition between interface minimization and baroclinic instability}

\begin{figure}
\begin{center}
\includegraphics[width=\textwidth]{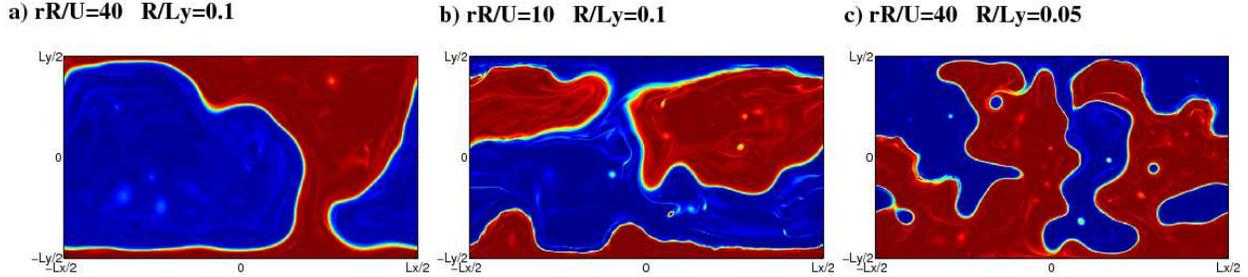}
\end{center}
\caption{Typical snapshots of the potential vorticity field a) in the reference case b) when bottom friction coefficient $rR/U$ is decreased c) when the Rossby radius  $R/L_y$ is decreased.  The typical  size of the potential vorticity blobs decreases from a to c and the interface perimeter increases from a to c.  \label{fig:blobs}}
\end{figure}

We show on Fig. \ref{fig:q_psi}-d a case where two simply connected regions of homogenized potential vorticity are formed. When bottom friction is decreased from  $rR/U =40$ to $rR/U=2$, we see on the histograms of potential vorticity levels Fig. \ref{fig:homog} that the global potential vorticity distribution still evolves to  a state characterized by a double delta function. However, a snapshot of the potential vorticity field Fig. \ref{fig:blobs}-b reveals that when $rR/U$ is decreased, the two peaks in the potential vorticity distribution  are associated with several unconnected blobs of regions with homogenized potential vorticity. The typical size of potential vorticity blobs decreases with lower bottom friction, while the total interface perimeter increases with lower bottom friction.  
%This last effect is clearly visible on Fig. \ref{fig:perimeter} were the temporal evolution of the interface perimeter is plotted for various values of the bottom friction coefficient $rR/U$. 
%To interpret the result, we consider an initial straight interface between two regions of homogenized potential vorticity, and a perturbation of this interface at scale $L_{flow} \gg R$.  On the one hand, the  inviscid  dynamics of the 1-1/2 layer model would tend to smooth out the distortion of the blob. Let us assume that this occurs at the time scale of the linear propagation of a wave, which gives a time scale $t_{relax}=L_flow U$, where we have assumed U_{jet}=U. On the other hand, the baroclinic instability of the jet tends to  increases the perimeter of the ribbons at a time scale $t_{diss}\sim rR^2/U^2$. If both time scales are of the same order then $L_{flow} \sim rR^2/U^3$. the typical length scale of the flow, the time $t_{turb}$ associated with turbulent  tee get from the inviscid potential vorticity equation the estimate $\psi_1/(\deltaR^2 t_{turb}) \sim \psi_1^2/L_{flow}^4$, which yields $t_{turb}=$The time scale for the blob agglomeration is independent of bottom friction. 
Similarly, we observed that decreasing the ratio $R/L_y$ for a given value of the bottom friction coefficient lead to an increase of the interface perimeters between region of homogenized potential vorticity, and to favor the detachment of isolated blobs of homogenized potential vorticity, see Fig. \ref{fig:blobs}-c.

%\begin{figure}
%\begin{center}
%\includegraphics[width=0.7\textwidth]{figures_ribbons/perimeter}
%\end{center}
%\caption{ Temporal evolution of a rough  estimate of the interface length between the regions of homogenized potential vorticity, for different values of the bottom friction coefficient $rR/U$. Time is adimensionalized by $t_{diss}=rRL_y/(\delta^{1/2} U^2)$ for each run in order to visualize all the runs on the same plot. 
 %\label{fig:perimeter}}
%\end{figure}

We interpret these observations by noting first  that destabilization of the ribbons occurs at a time scale of the energy decay controlled by the baroclinic instability, and given by $t_{diss} \sim (rRL_y/U^2)$ according to the large friction limit of Eq. (\ref{eq:t_diss_general}-\ref{eq:t_diss_F}). By contrast, the tendency of the 1-1/2 quasi-geostrophic dynamics to form simply connected regions of homogenized potential vorticity with minimal interface occurs at a time scale $t_{relax}$ independent from bottom friction parameter $rR/U$.  To estimate $t_{relax}$, we assume first that the flow is composed of two "phases" characterized by different values of potential vorticity, but that there are  several blobs associated with each phase (like bubbles in liquid water).  The potential vorticity jump between these two phases can be estimated to be initially  $\mathcal{Q}_1 \sim UL_y/\delta R^2$, which corresponds to stream function variations $\psi_1\sim U L_y$. Let us introduce $L_{flow}$ the typical length scale of a blob of homogenized potential vorticity. Then, assuming $L_{flow}\gg \delta^{1/2}R$ and $L_{flow} \ll L_y$ , and using the fact that the dynamics of the large scale flow  is given at lowest order by the planetary quasi-geostrophic model Eq. (\ref{eq:planetary-qg}), we obtain $\partial_t \psi_1/(\delta R^2 ) \sim J(\nabla^2 \psi_1,\psi_1)$ , which gives  $ UL_y/( t_{relax}\delta R^2) \sim (UL_y)^2/L_{flow}^4$.  A quasi-stationary state is reached when the relaxation time scale is of the order of  the baroclinic instability time scale ($t_{relax}\sim t_{diss}$), which  yields  
\begin{equation}
L_{flow} \sim L_y  \left(\frac{rR}{U}\right)^{1/4}  \left(\frac{R}{L_y}\right)^{1/2} \delta^{1/8}. \label{eq:Lflow}
\end{equation}
 The validity of scaling requires a scale separations that was not clear in our simulations (the potential vorticity blobs are not much smaller than the domain scale on Fig. \ref{fig:blobs}).  However, this naive scaling allows to interpret qualitatively our numerical results. The main point is that decreasing the bottom friction or the Rossby radius of deformation corresponds to a decrease of the typical size of $L_{flow}$ isolated blobs of potential vorticity, which means  an increase of the number of isolated blobs (since the goal area of a given phase is fixed), and therefore an increase of the total interface perimeter. We also note that the exponent $1/4$ means that variations of $L_{flow}$ are very weak when bottom friction is changed over one or two decades such as in our simulations.   Finally, we note that the length scale $L_{flow}$ for the homogenized potential vorticity blobs  can be interpreted as the scale of the  available potential energy field, and that  our scaling Eq. \ref{eq:Lflow}  is in very good agreement with the variations of the  potential energy centroids when bottom friction is varied in numerical simulation by \citet{arbicflierl2004} (figure 9-a of their paper). 
%PSI1/R^2 = Lflow^4

\subsubsection{Multiple jets}

\begin{figure}
\begin{center}
\includegraphics[width=\textwidth]{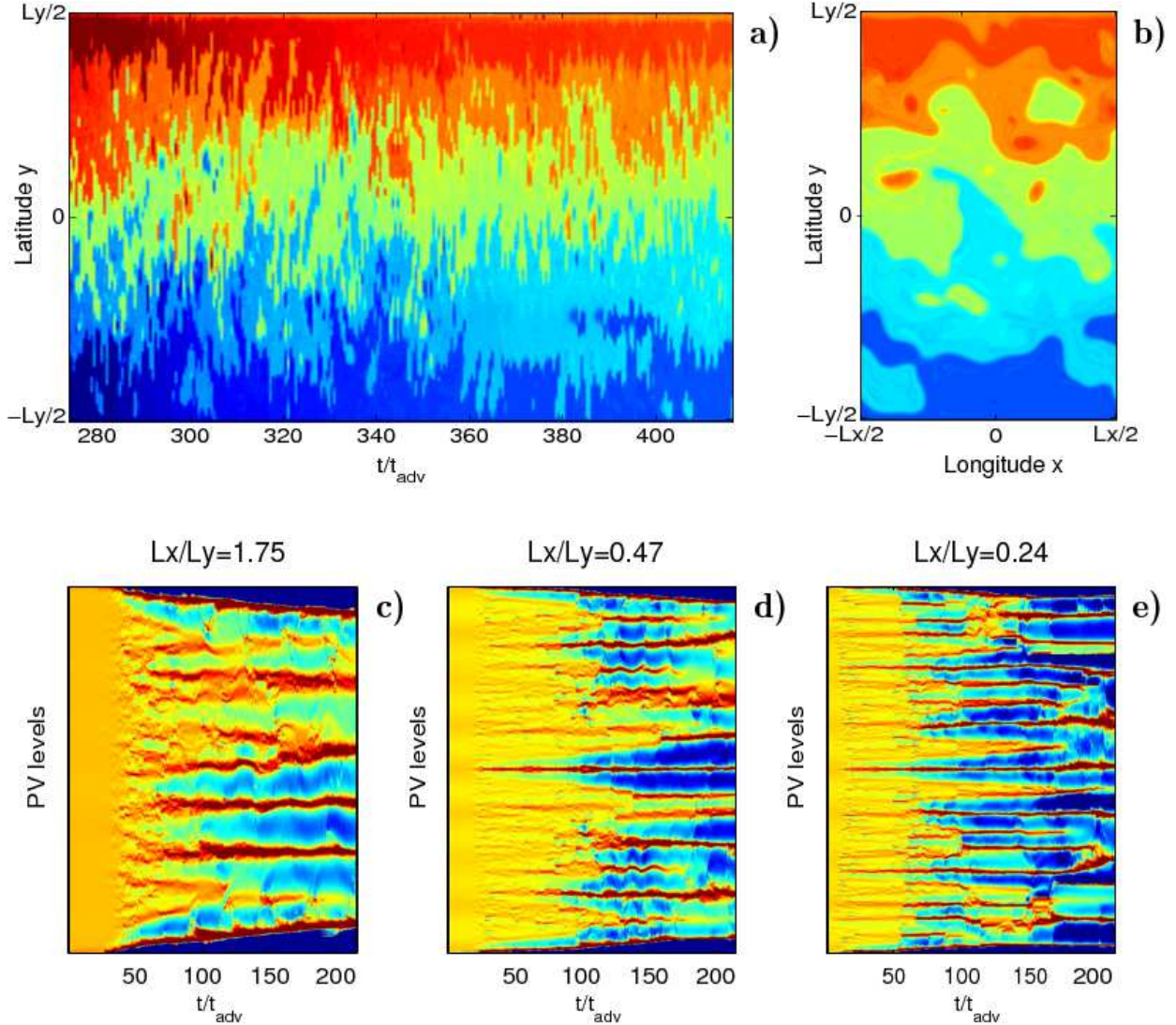}
\end{center}
\caption{ Multiple jets as a transient regime towards complete homogenization. a) Hovm\"oller diagrams of a meridional slice of the potential vorticity field in the upper layer, time unit is $t_{adv}=L_y/U$. b) Typical snapshot of the potential vorticity field in  the upper layer. c), d) and e)  Evolution of the global distribution of potential vorticity levels in the upper layer for different values of  the domain aspect ratio, $L_x/L_y=$1.75, 0.47 and 0.24 respectively.
 \label{fig:multiple}}
\end{figure}

We see on Fig. \ref{fig:historibbon}-b that there is a transient regime with multiple peaks in the global  potential vorticity distribution. These transient states correspond initially   to multiple regions  of homogenized potential vorticity.  We found that in the ribbon regime, the number of long lasting multiple regions of homogenized potential vorticity increased: i/ when the domain aspect ratio $L_x/L_y$ was decreased, ii/ when bottom friction $rR/U$ was increased and iii/ when the parameter $R/L_y$ was decreased. In addition, when the parameter $R/L_y$ was sufficiently small, the regions of homogenized potential vorticity are initially organized into zonal bands with east jet at their interface, which is reminiscent of potential vorticity staircases~\citep{dritschelmcintyre2008}. We in show Fig \ref{fig:multiple}-a,b an example of such long lasting multiple zonal bands of potential vorticity. In addition,
Fig \ref{fig:multiple}-c,d,e show how the number of regions of homogenized potential vorticity  increases with smaller domain aspect ratio $L_x/L_y$. It is not clear wether the dynamics would eventually form only two regions of homogenized potential vorticity, or if more than two regions of homogenized potential vorticity could last for ever. One may interpret qualitatively the emergence of these zonal potential vorticity staircases by  noticing that once a jet is formed between two regions of homogenized potential vorticity, it acts as a strong mixing barrier between the two adjacent regions, which may prevent further mixing with other regions of homogenized potential vorticity.  We note that in our case there is no beta effect. The zonal organization of the potential vorticity field only reflects the structure of the imposed mean flow, which induces an effective beta effect in the upper layer.  

The existence of long lived multiple eastward jets provides a route towards potential vorticity homogenization that sustains a total eastward transport of the order of the transport of the imposed mean flow. This contrasts with the low or intermediate bottom friction case where the rapid  decrease of the total potential energy (over a time $t_{adv}=L_y/U$)  is accompanied with a rapid  decrease of the total zonal transport. In that respect we find that increasing bottom friction leads to an increasing zonal transport in the regime where multiple jets are allowed. Increasing transport associated with increasing bottom friction  was reported in the context idealized simulations of the antarctic circumpolar circulation~\citep{nadeaustraub2012}, but this effect was due to the presence of bottom topography which is absent in our simulations. %

\section{Conclusion} \label{sec:conclude}

We have presented numerical simulations for the non-linear equilibration of a two-layer quasi-geostrophic flow in a channel in the presence of a baroclinically unstable imposed mean flow $U$ in the upper layer with particular attention to the role of bottom friction. For any non zero value of the bottom friction coefficient, $r$, the dynamics attempts to homogenize the potential vorticity field, including any large scale gradient due to imposed mean flow, as might expected from classical theories of geostrophic turbulence~\citep{RhinesYoung82}. This leads eventually to a perturbed flow that annihilates the imposed mean flow.  However, the  route toward complete homogenization depends strongly on the bottom friction coefficient.  

When the bottom friction is weak ($r \ll U/R$),  the perturbation self-organizes  at the domain scale into a quasi-barotropic large scale structure (see movie 1 in supplementary materials), which is then weakly dissipated on a time scale inversely proportional to the bottom friction coefficient, $t_{diss}\sim 1/r$. We interpret this large-scale  quasi-stationary flow as a statistical equilibrium state of the Miller-Robert-Sommeria (MRS) theory.

When  the bottom friction has a medium value --- meaning that its time scale is of the order of the inviscid baroclinic instability time scale ($r \sim U/R$) --- bottom friction  precludes an inverse kinetic energy cascade close to the injection length scale (which is of the order of the Rossby radius deformation $R$) and the dynamics is well described by a gas of isolated vortices of size $R$ mixing the background potential vorticity field at the advection time scale $t_{diss} \sim L_y/U$  (see movie 2 in supplementary materials).

When the bottom friction coefficient is high ($r \gg U/R$), the ratio between the lower layer kinetic energy and the upper layer kinetic energy scales as $(rR/U)^2$ and the dynamics is well described at lowest order by a 1-1/2 layer quasi-geostrophic model. We observed the spontaneous emergence of meandering ribbons corresponding to strong jets of width given by the Rossby radius of deformation of the upper layer, and separating regions of homogenized potential vorticity   (see movie 3 in supplementary materials). We used statistical mechanics arguments as well as cascade phenomenology to interpret these results. We described a competition between the inviscid 1-1/2 quasi-geostrophic dynamics that tends to form only two regions of homogenized potential vorticity with a minimal interface between them, and baroclinic instability of the ribbons that tend to increase the interface perimeter. This last route towards potential vorticity homogenization is rather spectacular: the potential vorticity jump between the two regions of homogenized potential vorticity decreases slowly with time, due to the intermittent breaking of the ribbons at their interface. This process occurs at a time scale given by baroclinic instability that scales linearly with the bottom friction coefficient $t_{diss} \sim rRL_y/U^2$. Remarkably,  the interface between the homogenized regions of potential vorticity remains sharp (i.e. much smaller than the Rossby radius of deformation) throughout this evolution towards a single, fully homogenized potential vorticity field. 

%We also performed numerical simulations in a doubly periodic domain with an imposed mean flow, in which case the dynamics reaches a statistically stationary state. We found that the time $t_{diss} \sim rRL_y/U^2 $ is an intrinsic time scale for the variability of the available potential energy in that case.  
 Using cascade phenomenology, and generalizing the arguments by \citet{heldlarichev1996},  \citet{arbicflierl2004} proposed scalings for the horizontal scale and the vertical structure of the dynamics in the large friction regime. Here we obtained rather different scalings, but with similar qualitative meaning, by assuming that the flow structures resulted from the competition between baroclinic instability and a tendency to reach a MRS equilibrium state in both the weak  and the large bottom friction limit. We believe that the cascade arguments are more suited to intermediate bottom friction, for which there is a scale separation between the large scale flow and the perturbed flow.  A key novelty of our work is to relate the emergence of the ribbons with existing statistical predictions for the 1-1/2 layer quasi-geostrophic model. In particular, we show for the first time numerical evidence that when the Rossby radius of deformation is much smaller than the domain scale, the dynamics  attract the system towards a quasi-stationary state characterized by a tanh-like relation between potential vorticity and stream function, even if the initial potential vorticity distribution is not already made of several regions with homogenized potential vorticity.  We note that in our case the presence two layers was essential to observe large regions of homogenized potential vorticity, even if the  dynamics is described at lowest order by a 1-1/2 layer quasi-geostrophic flow. Indeed, the presence of the bottom layer allows for baroclinic instability of the ribbons, which favors stirring of the upper layer potential vorticity field in the whole flow domain. By contrast, once a ribbon emerges in a freely evolving 1-1/2 quasi-geostrophic flow, it acts as a mixing barriers that prevent further exchanges between adjacent regions of homogenized potential vorticity.
 
Our work was set in a channel geometry in which case the global distribution of a suitably defined potential vorticity field is conserved in the absence of small scale dissipation. This allows us to use of statistical mechanics arguments and reinterpret the results obtained in previous work in doubly periodic geometry.  Thus, in the large bottom friction limit,  the dissipation time $t_{diss} \sim rRL_y/U^2 $ can be interpreted as an intrinsic time scale for the variability of the available potential energy in a statistically steady state.
It is also interesting to compare our results with those of \citet{esler2008,willcocks2012nonlinear} who considered the free evolution of an surface intensified zonal jet localized at the center of a channel. In their case, the instability is localized around the jet, and potential vorticity stirring occurs only within this central region. Statistical mechanics predictions fail in this case to predict the large scale flow structure since the dynamics does only explore a restricted part of the phase space. By contrast, in our simulations, the initial instability and its subsequent turbulent evolution takes place in the whole domain, which induces potential vorticity stirring everywhere, excepted when multiple jets occur. 
 
To conclude, this study shows that large bottom friction induces the condensation of the kinetic energy into quasi-stationary ribbons and the concomitant condensation of potential energy at large scale. Perhaps paradoxically increasing the bottom friction considerably slows down the loss of energy from the potential energy reservoir associated with the large scale flow. 

The regime for ribbons turbulence requires bottom friction coefficient which are too high for a direct application to oceanic flows. However, other physical mechanism than bottom drag may be able to remove energy from the lower layer, which would mimmic the effect of high bottom friction. For instance \citet{lacasce2000}  showed in the framework of freely solving two-layer quasi-geostrophic turbulence over a slope that topographic Rossby waves generated in some location remove the energy to other locations, where it eventually is dissipated by bottom drag. This effect may me interpreted as an enhanced bottom friction in the region where the topographic wave is generated. 
 
 Further work will be needed to extend these results to continuously stratified fluids because in that case other effects can significantly change the properties of the vertical structure of the eddies, see \citet{smith2002,RouletJPO2012} for the forced dissipated case, and \citet{smith2001} for the freely evolving case.  In particular, \citet{smith2001,fuflierl1980}  did show that in the presence of surface intensified stratification,  and without bottom friction, there is a fast time scale associated with  energy transfers toward the first baroclinic mode. This energy eventually condense into the barotropic mode, but with a much larger time scale.  The beta effect may also have several consequences: it is known to favor barotropization~\citep{VenailleVallisGriffies}, and to favor the arrangement of regions of homogenized potential vorticity into zonal bands. %Bottom topography may also play an important role, either by favoring the formation of bottom in tensified flow through turbulent stirring \icte{venaille}, or by allowing for Rossby topographic waves that propagate the energy away \cite{lacsce2000}.

Finally, we note that we have here considered only an imposed mean flow and it would be useful to generalize to a more realistic forcing by considering a surface wind stress or relaxation towards a prescribed unstable flow.  We conjecture that our estimate for the dissipation time $t_{diss}$ will still play an important role to describe low frequency, internal variability of the system, with the only difference that the estimate of the large scale velocity $U$ will have to be related to the forcing case by case.

\paragraph{\itshape Acknowledgments}
%The authors warmly thank  F. Bouchet and blablabla for useful discussions. AV  was funded by NOAA??and ANR LORIS??, and CNRS  for a large part of this project, LPN was funded by GRANTS 

\section{Appendix A Barotropization in the weak bottom friction limit}

The aim of this appendix is to give a phenomenological argument for barotropization when $R \ll L_y$ or $\delta \ll 1$, with $L_x \sim L_y$. The argument is based on the fact that turbulence leads to a rearrangement of the initial potential field in each layer, with a constant total energy $E_{tot}$.  
%We recall that the initial potential vorticity fields in the upper and lower layers are respectively  $q_1^0=Uy /\delta R^2$ and $q_2^0=-U y /(1-\delta) R^2$. 

 The global distribution of potential vorticity levels in both layers are conserved when there is neither small scale dissipation nor bottom friction.  
 
 Let us call $ \mathcal{Q}_1$ the typical variations of the potential vorticity field in the upper layer after turbulent rearrangement of the initial field  $q_1^0=U y /(\delta R^2)$. We see Eq.  (\ref{eq:barotrope})  that  typical variations of the barotropic streamfunction are given by   $\psi_t \sim \delta  L_y^2\mathcal{Q}_1$, where we anticipate that the typical length scale of flow structures in this regime is given by the domain size $L_y$. We also see from Eq. (\ref{eq:barocline}) that typical variations of the baroclinic streamfunction  are  $(\psi_c -Uy )\sim  \delta (1-\delta ) R^2 \mathcal{Q}_1$ over a length $(\delta  (1-\delta ))^{1/2} R$ when $(\delta  (1-\delta ))^{1/2} R \ll L_y $.  With these estimates, and anticipating that $\psi_t \gg Uy$, we find the following scalings for the different components of the energy of the perturbed flow introduced Eq. (\ref{define_Etot}):
\begin{equation}
KE_{tot,t} \sim  \mathcal{Q}_1^2 L_y^4 \ ,\quad  KE_{tot,c} \sim   \mathcal{Q}_1^2 \delta (1-\delta) {R^3} L_y  \ ,  \quad APE_{tot,c} \sim  \mathcal{Q}_1^2 \delta^2 R^2 L_y^2 \ .\label{eq:Energy}
\end{equation}
Clearly, the total energy  $E_{tot}=KE_{tot,t} +KE_{tot,c}+APE_{tot,c}$  is dominated its barotropic component  $KE_{tot,t}$ when $\delta \ll 1$ or $R\ll L_y$.  Since the barotropic dynamics  leads to an inverse kinetic energy cascade, our hypothesis that  $L_y$  is a typical scale of the flow is self-consistent.  Using the estimate of the initial energy  $E^0_{tot}\sim APE_{tot}^0 \sim U^2L_y^4/R^2$, and using the fact that this energy is fully transferred into the barotropic mode after turbulent rearrangement, we get  $KE_{tot,t} \sim U^2L_y^4/R^2$. Using Eq. (\ref{eq:Energy}),  this estimate yields $\mathcal{Q}_1 \sim  U/R$. Consequently, the order of magnitude for the barotropic velocity is  $U_t\sim UL_y/R$, which is consistent with the hypothesis that the barotropic  flow is dominated by the perturbed flow ($\psi_t\gg \delta Uy$). We conclude that the total flow is dominated by the barotropic component of the perturbed flow when $\delta \ll 1$ or $R\ll L_y$ after turbulent rearrangement of the potential vorticity field. 
%Let us come back to the energy budget Eq. (\ref{eq:Energy-dynamics}). When $r=0$, the second term of the right hand side of the energy budget is zero. In addition, we have just seen that $\psi_1=\psi_2$ at lowest order since the dynamics is dominated by the barotropic mode. Then the forcing term  $(1/R)\int_{\mathcal{D}}\mathrm{d}x\mathrm{d}y\psi_1\partial_x \psi_2$ vanish at lowest order. 

%\bibliographystyle{pf}

\bibliography{total}

\end{document}